\documentclass[aps,prd,twocolumn,showpacs]{revtex4} 

\usepackage{graphicx}

\font\FermiSmallfont=cmssq8 scaled 1200

\def\LANLppthead#1#2#3#4#5#6{
\null 
\begin{center}\vskip -1.0truein{\hbox to 7.5truein {
\vbox to 1.1in {\vfill \FermiSmallfont
              \hbox{#1}
              \hbox{#2}
              \hbox{#3}
              \vfill}
\hfill
\vbox to 1.1in {\vfill \FermiSmallfont
              \hbox{#4}
              \hbox{#5}
              \hbox{#6}
              \vfill}
}}\vskip-0.0truein\end{center}}

\begin{document}
\LANLppthead {LA-UR-04-2260}{KRL-MAP-303}{DESY-04-202}{FNAL-PUB-04-263-A}{NSF-KITP-04-113}{astro-ph/0410175}

\title{Cosmological Lepton Asymmetry, Primordial Nucleosynthesis\\ and Sterile Neutrinos}
\author{Kevork Abazajian$^{1,2,3}$, Nicole F.\ Bell$^{2,3,4}$, George M.\ Fuller$^{3,5}$, Yvonne Y.\ Y.\ Wong$^{3,6}$}
\affiliation{$^1$Theoretical Division, MS B285, Los Alamos National Laboratory,
 Los Alamos NM 87545\\ 
$^2$Theoretical Astrophysics, Fermi National Accelerator
 Laboratory, Batavia IL 60510\\ 
$^3$Kavli Institute for Theoretical Physics, University of California, Santa
Barbara, CA 93106\\
$^4$Kellogg Radiation Laboratory, California Institute of Technology,
Pasadena, CA 91125\\
$^5$Department of Physics, University of California, San Diego, La Jolla, CA
92093-0319\\
$^6$ Deutsches Elektronen-Synchrotron DESY, D-22607 Hamburg, Germany}
\date{August 19, 2005}

\begin{abstract}
We study post weak decoupling coherent active-sterile and active-active
matter-enhanced neutrino flavor transformation in the early universe.  We show
that flavor conversion efficiency at Mikheyev-Smirnov-Wolfenstein resonances is
likely to be high (adiabatic evolution) for relevant neutrino parameters and
energies.  However, we point out that these resonances cannot sweep smoothly
and continuously with the expansion of the universe.  We show how neutrino
flavor conversion in this way can leave both the active and sterile neutrinos
with non-thermal energy spectra, and how, in turn, these distorted energy
spectra can affect the neutron-to-proton ratio, primordial nucleosynthesis, and
cosmological mass/closure constraints on sterile neutrinos.  We demonstrate
that the existence of a light sterile neutrino which mixes with active
neutrinos
 can change fundamentally the relationship between the cosmological
lepton numbers and the primordial nucleosynthesis $^4${He} yield.
\end{abstract}
\maketitle

\section{Introduction}
\label{sec-1}
If light sterile neutrinos exist we will be forced to re-think the role of the
weak interaction in the early universe, primordial nucleosynthesis, and
cosmology.  Though light sterile neutrinos which mix with active neutrinos long
have been a subject of theoretical speculation, the LSND and KARMEN experiments
\cite{LSND} gave rise to particular interest in the mass-squared difference
range $0.2\,{\rm eV}^2<\delta m^2_{\rm as}<100\,{\rm eV}^2$.  Here we study for
this range of $\delta m^2_{\rm as}$ the cosmological lepton number-driven
conversion of active neutrinos, $\nu_\alpha$ (and/or $\bar\nu_\alpha$) with
$\alpha={\rm e},\mu,\tau$, to a singlet, \lq\lq sterile\rq\rq\ neutrino species
$\nu_s$ (or $\bar\nu_s$) in the epoch of the early universe after decoupling of
the weak interactions, when neutrino spectral distortions are likely to
persist.

A positive signal in the on-going mini-BooNE experiment \cite{miniB}, {\it
  i.e.}, confirming the interpretation of the LSND result in terms of vacuum
neutrino mixing, sets up an immediate crisis in neutrino physics. Such a
result, when combined with the already well established evidence for neutrino
mixing at mass-squared differences associated with the atmospheric ($\delta m^2
\sim 3\times{10}^{-3}\,{\rm eV}^2$) and solar neutrino ($\delta m^2 \sim
7\times{10}^{-5}\,{\rm eV}^2$) anomalies, would suggest the existence of three
independent neutrino mass-squared differences which would, in turn, require
four neutrino species. Given the $Z^0$-width limit on the number of flavors of
neutrinos with standard weak interactions (3), a fourth neutrino would have to
be \lq\lq sterile,\rq\rq\ with sub-weak interaction strength, {\it e.g.},
perhaps an SU(2) singlet. The only alternative to this line of reasoning and to
this conclusion is the possibility of CPT violation~\cite{cpt1}.  However,
there is no consistency of the neutrino oscillation data with a CPT-violating
three-neutrino model at a 3-$\sigma$ level~\cite{cpt2}.

Hand in hand with this particle physics dilemma, evidence for a singlet
neutrino that mixes with active neutrinos in this mass-squared range also
confronts cosmology with a curious and vexing problem. In the standard
cosmological model with zero or near-zero net lepton numbers one would expect
that matter-supressed neutrino oscillations in the channel $\nu_\alpha
\rightleftharpoons\nu_s$ or in $\bar\nu_\alpha \rightleftharpoons \bar\nu_s$
(where $\alpha=e,\mu,\tau$) proceeding in the regime above weak interaction
decoupling ($T > 3\,{\rm MeV}$), would efficiently populate seas of singlet
neutrinos~\cite{aba}. The significant additional energy density in these
sterile neutrino seas would engender a faster expansion rate for the universe
and a consequently higher temperature for Weak Freeze-Out (where the initial
isospin of the universe, the neutron-to-proton ratio is set).  A higher Weak
Freeze-Out temperature would result in more neutrons and, hence, a higher yield
of {$^4${He}}.

A higher predicted abundance of $^4$He {\it arguably} may be in
conflict (or is close to being in conflict) with the observationally
inferred upper limit on the primordial helium abundance.  Depending on
the helium abundance inferred from compact blue galaxies, an increase
in the predicted Big Bang Nucleosynthesis (BBN) $^4$He yield may or
may not be disfavored \cite{aba,cyburt}.  However, the primordial
helium abundance is notoriously difficult to extract from the
observational data and recent studies point to a fair range for the
observationally inferred primordial mass helium fraction: $23\%$ to
$26\%$ \cite{OS}. The upper limit of this range is provocatively close
to the standard BBN $^4$He mass fraction yield prediction, $24.85\pm
0.05\%$, as computed with the deuterium-determined or CMB (Cosmic
Microwave Background) anisotropy-determined baryon density.

Additionally, it has been suggested \cite{murayama} that a fully populated sea
of sterile neutrinos and antineutrinos with rest masses $\sim \sqrt{\delta
  m^2_{\rm as}}$ could be in conflict with neutrino mass bounds derived from
CMB anisotropy limits and large scale structure considerations
\cite{WMAP}. There is a recent analysis of constraints from measurements of
galaxy bias stemming from galaxy-galaxy lensing and the inferred linear matter
power spectrum derived from the Lyman alpha forest in the Sloan Digital Sky
Survey (SDSS) \cite{Sloan}.  This analysis specifically considers a so-called
\lq\lq 3+1\rq\rq\ neutrino mass hierarchy, {\it i.e.,} the scheme which is
appropriate for constraining sterile neutrinos. The neutrino mass constraint so
derived is somewhat less stringent than constraints in schemes with three
neutrinos with degenerate masses. However, the central conclusions of
Ref. \cite{murayama} survive.

Should we someday be confronted with a positive indication of neutrino flavor
mixing with mass-squared scale consistent with the range for $\delta m^2_{\rm
  as}$, we will have a problem that would call for modification either of our
notions of basic neutrino physics or of the standard cosmological model.  There
have been a number of ways proposed to get out of these cosmological
difficulties.  For instance, if neutrinos do not acquire mass until after the
BBN epoch, as may occur via a late-time phase transition
\cite{flypaper,latetime}, the singlet states will not be populated via
oscillations during the BBN era. In addition to schemes involving the epoch of
neutrino mass generation \cite{Chacko:2003dt}, annihilation \cite{neutrinoless}
or decay \cite{latetime} of the singlet neutrinos (when $m_s \sim T$) may
alleviate or avoid the CMB and large scale structure constraints.  However,
chief among the mechanisms proposed to escape the cosmological difficulties
associated with singlet neutrinos, is the the invocation of a significant net
lepton number \cite{FV}.

The idea is that the net lepton number gives active neutrinos larger effective
masses in medium in the early universe, thereby driving them further
off-resonance in the epoch prior to weak decoupling ({\it i.e.}, $T>3\,{\rm
  MeV}$) and reducing their effective matter mixing angles with the singlet
neutrino. In turn, smaller effective matter mixing angles would imply a
suppressed production of singlet neutrinos and, hence, a reduced population of
the singlet neutrino sea.  This lepton number-induced suppression of
active-sterile mixing at high temperature is why we assume here that there is
no initial population of the sterile neutrino sea.

The lepton number residing in the sea of $\nu_\alpha$ and $\bar\nu_\alpha$
neutrinos ($\alpha=e,\mu,\tau$) is defined in analogy to the baryon number
$\eta \equiv (n_b-n_{\bar b})/n_\gamma$,
\begin{equation}
L_{\nu_{\alpha}}= {{n_{\nu_\alpha}-n_{\bar\nu_\alpha}}\over{n_\gamma}}
\label{lepno}
\end{equation}
where $n_\gamma = \left(2\zeta(3)/\pi^2\right) T^3_\gamma$ is the proper photon
number density at temperature $T_\gamma$, and where $n_{\nu_\alpha}$ and
$n_{\bar\nu_\alpha}$ are the number densities of $\nu_\alpha$ and
$\bar\nu_\alpha$ neutrinos, respectively, at this epoch. After the epoch of
$e^\pm$ annihilation the baryon number is $\eta \approx 6\times10^{-10}$,
whereas at earlier epochs it is roughly two and half times larger. We will
consider here net lepton numbers which are vastly larger than $\eta$, so its
precise value is of no consequence for our results.

We can insure that the effective matter mixing angles for the oscillation
channel $\nu_\alpha \rightleftharpoons\nu_s$ (or $\bar\nu_\alpha
\rightleftharpoons\bar\nu_s$) are sufficiently small to suppress singlet
neutrino production if the Mikheyev-Smirnov-Wolfenstein (MSW) \cite{MSW}
resonance temperature is less than the weak decoupling temperature, $T_{\rm
  res} < T_{\rm dec}$. This implies that the lepton number associated with any
of the active neutrino flavors should satisfy,
\begin{equation}
L > {{10^{-3}}\over{\epsilon}} \left({{2}\over{N_{\rm degen}}}\right) {\left({{3\,{\rm MeV}}\over{T_{\rm dec}}}\right)}^4 \left({{\delta m^2_{\rm as} \cos2\theta}\over{1\,{\rm eV}^2}}\right)
\label{suppress}
\end{equation}
where $\theta$ is the vacuum mixing angle characteristic of
$\nu_\alpha\rightleftharpoons\nu_s$ oscillations, $N_{\rm degen}$ is the number
of neutrino species possessing this lepton number, and where $\epsilon \equiv
E_\nu/T$. For neutrinos with typical energies in the early universe ({\it
  i.e.}, $\epsilon \sim 1$), suppression of singlet neutrino production would
require lepton numbers ranging from $L > {10}^{-4}$ for $\delta m^2_{as}
=0.2\,{\rm eV}^2$ to $L > 5\times{10}^{-3}$ for $\delta m^2_{as} =10\,{\rm
  eV}^2$. Current limits on lepton numbers are $\vert {L_{\nu_\alpha}}\vert <
0.1$ \cite{abb} (and possibly even weaker by a factor of two or so if allowance
is made for another source of extra energy density in the early universe
\cite{barger,Kneller:2001cd}). Therefore, this avenue for escape from the
sterile neutrino conundrum appears to be allowed, albeit at the cost of a huge
disparity between the lepton and baryon numbers.

However, this argument overlooks an important point. Though the large
lepton number supresses the effective matter mixing angle for
$\nu_\alpha\rightleftharpoons\nu_s$ during the epoch of the early
universe where active neutrinos are thermally coupled ($T>T_{\rm
  dec}$), it can cause coherent matter-enhancement of this channel at
lower temperatures where the active neutrinos rarely scatter and are
effectively decoupled. Resonant MSW transformation of active neutrinos
to singlets in the channel $\nu_\alpha\rightleftharpoons\nu_s$ is,
however, self limiting. This is because as the universe expands and
the resonance sweeps from low toward higher neutrino energy, the
conversion of $\nu_\alpha$'s decreases the lepton number which, in
turn, causes the resonance sweep rate to increase, eventually causing
neutrinos to evolve non-adiabatically through resonance and so causing
flavor transformation to cease.

At issue then is how many active neutrinos can be converted to sterile
neutrinos prior to or during the epoch where the neutron-to-proton ratio is set
(\lq\lq Weak Freeze-Out\rq\rq). If there is a significant conversion, the
resultant non-thermal active neutrino energy spectra can cause an increase or
decrease (if $\bar\nu_\alpha \rightleftharpoons \bar\nu_s$ is enhanced) in the
$^4$He yield and call into question the viability of invoking a large net
lepton number to reconcile neutrino physics and BBN. Other but related aspects
of transformation-induced nonthermal neutrino spectra effects on primordial
nucleosynthesis have been studied in Ref.s~\cite{Abazajian:1999yr,
  Kirilova:2001ab}. Likewise, previous studies have considered other aspects of
the relationship between sterile neutrinos and BBN \cite{new}, as well as
constraints on sterile neutrinos without a primordial lepton number
\cite{Cirelli:2004cz}.  In any case, non-thermal energy distribution functions
for $\nu_e$ and/or $\bar\nu_e$ change the relationship between the BBN $^4${He}
yield and the neutrino chemical potentials.

In section II we discuss the physics of active-sterile neutrino flavor
transformation in the early universe and point out a key issue in how the MSW
resonance sweeps through the neutrino energy distribution functions as the
universe expands. The generally high adiabaticity of neutrino flavor evolution
is also pointed out in this section.  Simultaneous active-active and
active-sterile neutrino flavor conversion, and \lq\lq
synchronization\rq\rq\ are also discussed in this section.  Possible
multi-neutrino mass level crossing scenarios in the early universe are
discussed in this section. Sterile neutrino contributions to closure,
constraints on this from large scale structure and Cosmic Microwave Background
radiation considerations, as well as other sterile neutrino sea population
constraints are examined in section III.  In section IV we describe how
distorted $\nu_e$ and/or $\bar\nu_e$ distribution functions impact the rates of
the lepton capture reactions that determine the neutron-to-proton ratio and the
$^4$He yield in BBN. This is then applied in various initial lepton number and
neutrino conversion scenarios. Finally, in section V we give conclusions and
speculations regarding the neutrino mass and cosmological lepton number
insights that would follow in the wake of an experimental signature for a large
neutrino mass-squared difference of order the range given for $\delta
m^2_{as}$. Appendix A provides an exposition of the lepton capture rates on
free nucleons when, as appropriate, $\nu_e$ or $\bar\nu_e$ energy distribution
functions are zero up to some energy, and thermal/Fermi-Dirac at higher
energies.

\section{Coherent Neutrino Flavor Transformation in the Early Universe}

Coherent conversion of active neutrino species into singlets in the early
universe can occur through the usual MSW process, albeit in an exotic
setting. This process can be described simply: (1) an active neutrino (mostly
the light mass state in vacuum) forward scatters on particles in the plasma
and, if there is a net lepton and/or baryon number, will acquire a positive
effective mass; (2) an MSW resonance (mass level crossing, where in-medium
mixing is large) can occur when this effective mass is close to the mass
associated with the singlet (mostly the heavy mass state).  This process is
discussed in subsection~{ A}, but despite the simplicity of the physics behind
it, the neutrino energy dependence/history of MSW resonances in the early
universe can be quite complex, as shown in subsection~{B}.

The efficiency of flavor conversion at a mass level crossing
depends on the ratio of the resonance width (in time or
space) to the neutrino oscillation length. Efficient, adiabatic conversion
takes place only when this ratio is large, and in subsection~{C} we examine
this physics in detail for our particular problem.

Active-active matter-enhanced neutrino flavor conversion, discussed in
subsection~{D}, can occur simultaneously with active-sterile transformation in
the early universe. This can greatly complicate computing the history of the
neutrino distribution functions. Subsection~{E} below deals with the limit
where active-active conversion is efficient, while subsection~{ F} examines
flavor evolution in the limit where this conversion channel is inefficient.

\subsection{Neutrino Effective Masses and Level Crossings}

The forward charged and neutral current exchange Hamiltonians for the neutrinos
in the early universe are as follows (see, {\it e.g.,} \cite{Notzold:1987ik}):
\begin{equation}
H\left( \nu_s\right)=0
\label{hsterile}
\end{equation}
\begin{widetext}
\begin{equation}
H\left(\nu_e\right)={\sqrt{2}}G_{\rm F}{\left(n_e-{{1}\over{2}}n_n\right)}+{\sqrt{2}}G_{\rm F}{\left( 2\left(n_{\nu_e}-n_{\bar\nu_e}\right)+\left(n_{\nu_\mu}-n_{\bar\nu_\mu}\right)+\left(n_{\nu_\tau}-n_{\bar\nu_\tau}\right)\right)}
\label{hnue}
\end{equation}
\begin{equation}
H\left(\nu_\mu\right)={\sqrt{2}}G_{\rm F}{\left(-{{1}\over{2}}n_n\right)}+{\sqrt{2}}G_{\rm F}{\left( \left(n_{\nu_e}-n_{\bar\nu_e}\right)+2\left(n_{\nu_\mu}-n_{\bar\nu_\mu}\right)+\left(n_{\nu_\tau}-n_{\bar\nu_\tau}\right)\right)}
\label{hnumu}
\end{equation}
\begin{equation}
H\left(\nu_\tau\right)={\sqrt{2}}G_{\rm F}{\left(-{{1}\over{2}}n_n\right)}+{\sqrt{2}}G_{\rm F}{\left( \left(n_{\nu_e}-n_{\bar\nu_e}\right)+\left(n_{\nu_\mu}-n_{\bar\nu_\mu}\right)+2\left(n_{\nu_\tau}-n_{\bar\nu_\tau}\right)\right)}.
\label{hnutau}
\end{equation}
\end{widetext}
Here $n_e=n_{e^-}-n_{e^+}$ is the net number density of electrons,
$n_n=n_b-n_p$ is the number density of neutrons, and $n_b$ and $n_p$ are the
net number densities of baryons and protons, respectively. Charge neutrality
implies that the number density of protons is $n_p=n_e=n_b Y_e$. The net number
of electrons per baryon is $Y_e$. The baryon number density is $n_b\approx\eta
n_\gamma$, where the baryon-to-photon ratio $\eta$ is as defined above.

Weak Decoupling occurs when neutrino scattering becomes so slow that it can no
longer facilitate efficient energy exchange between the neutrino gas and the
plasma. For the low lepton numbers considered here, Weak Decoupling occurs
around temperature $T\sim 3\,{\rm MeV}$, though this decoupling process takes
place over a range in temperature of a few MeV.

Weak Freeze-Out occurs when the rates of the reactions that govern the ratio of
neutrons-to-protons ($n/p=1/Y_e-1$) fall below the expansion rate of the
universe. This is usually taken to be $T\approx 0.7\,{\rm MeV}$ for standard
cosmological parameters. However, below this temperature $Y_e$ continues to be
modified by lepton capture and/or free neutron decay as discussed below.

Note that for temperatures well above Weak Decoupling, we have $Y_e\approx
0.5$, {\it i.e.}, nearly equal numbers of neutrons and protons. We could then
approximate $n_e-{{1}\over{2}}n_n=n_b({{3}\over{2}}Y_e-{{1}\over{2}})\approx
{n_\gamma}\eta/4$, and $-{{1}\over{2}} n_n =n_b(Y_e/2-{{1}\over{2}})
\approx-n_\gamma\eta/4$. We will use this approximation in what follows even
for the epoch below weak decoupling where it is not numerically accurate.  This
will result in no loss of accuracy in the full calculation because we consider
large net lepton numbers, $L_{\nu_\alpha}\gg\eta$.

We can denote the weak potentials from neutrino-electron charged current
forward exchange scattering and neutrino-neutrino neutral current forward
exchange scattering as $A$ and $B$, respectively, with their sum being
\begin{equation}
A+B\approx{{2\sqrt{2}\zeta{\left(3\right)} G_{\rm F} T^3}\over{\pi^2}}{\left(
  {\cal{L}}\pm{{\eta}\over{4}}\right)},
\label{AB}
\end{equation}
where $G_{\rm F}$ is the Fermi constant, the Riemann Zeta function of argument
$3$ is $\zeta\left(3\right) \approx 1.20206$, and we take the plus sign for
transformation of $\nu_e$, and the minus sign for conversion of $\nu_\mu$
and/or $\nu_\tau$. (Here the plus sign is taken when we intend $A+B= H(\nu_e)$
and the minus sign is taken when $A+B=H(\nu_{\mu,\tau})$.)  A measure of the
lepton number which enters into the potential for the
$\nu_{\alpha}\rightleftharpoons\nu_s$ ($\alpha=e,\mu,\tau$) conversion channel
is
\begin{equation}
{\cal{L} } \equiv 2L_{\nu_\alpha}+\sum_{\beta\neq\alpha}L_{\nu_\beta}.
\label{Lpot}
\end{equation}
We will refer to this quantity as the \lq\lq potential lepton number.\rq\rq\ In
general this may be different for different channels
$\nu_\alpha\rightleftharpoons\nu_s$, even for a given set of lepton numbers
associated with each flavor.

Finally, since the early universe is at relatively high entropy per baryon, the
overall weak potential has a contribution from neutrino neutral current forward
scattering on a thermal lepton background. This thermal potential is
\begin{equation}
C\approx -r_{\alpha} G_{\rm F}^2 \epsilon T^5,
\label{thermal}
\end{equation}
where the neutrino energy divided by the temperature is $\epsilon \equiv
E_{\nu}/T$. For the conversion channel $\nu_e\rightleftharpoons\nu_s$, we
employ $r_e^0 \approx 79.34$, while for the channel
$\nu_{\mu,\tau}\rightleftharpoons\nu_s$, we use $r_{\mu,\tau}^0 \approx
22.22$. If the neutrinos have strictly thermal energy distribution functions,
then
\begin{equation}
r_\alpha \approx r_{\alpha}^0 \left[
{{F_2\left(\eta_{\nu_\alpha}\right)}\over{F_2\left( 0\right)}} +
{{F_2\left(\eta_{\bar\nu_\alpha}\right)}\over{F_2\left( 0\right)}} \right],
\label{ralpha}
\end{equation}
where the neutrino and antineutrino degeneracy parameters are
$\eta_{\nu_\alpha}$ and $\eta_{\bar\nu_\alpha}$, respectively, and the Fermi
integrals of order $2$ are defined below. The approximations $r_e\approx r_e^0$
and $r_{\mu,\tau}\approx r_{\mu,\tau}^0$ suffice for lepton numbers below the
conventional limits.

The total weak forward scattering potential is
\begin{equation}
V= A+B+C.
\label{totalV}\end{equation}
For the transformation channel $\nu_{\alpha}\rightleftharpoons\nu_s$, the
neutrino mass level crossing (MSW resonance) condition for a neutrino with
scaled energy $\epsilon$ is
\begin{equation}
{{{\delta m^2} \cos2\theta}\over{2\epsilon T}} = V,
\label{res}
\end{equation}
where $\delta m^2$ is the difference of the squares of the appropriate neutrino
mass eigenvalues and $\theta$ is the relevant {\it effective} two-by-two vacuum
mixing angle. Neglecting the light mass eigenvalue, the effective mass-squared
acquired by an electron neutrino from forward scattering on weak
charge-carrying targets in the early universe is
\begin{widetext}
\begin{equation}
m_{\rm eff}^2 \approx 2\epsilon V\approx \left(8.03\times{10}^{-12}\,{\rm
MeV}^2\right) \epsilon {\left( {\cal{L}} \pm \eta/4 \right)} {\left({{T}\over{\rm
MeV}}\right)}^4-\left(2.16\times{10}^{-20}\,{\rm MeV}^2\right) {\epsilon}^2
{\left({{T}\over{\rm MeV}}\right)}^6.
\label{meff}
\end{equation}
\end{widetext}
It is clear that we can negelect the second term (the thermal term $C$) in
Eq.~(\ref{totalV}) in the regime between Weak Decoupling and Weak Freeze-Out,
where $3\,{\rm MeV} > T> 0.7\,{\rm MeV}$. We also neglect the baryon/electron
term, $\pm \eta/4$.

At a given temperature, the scaled neutrino energy which is resonant is then 
\begin{equation}
{\epsilon}_{\rm res} = {{{\delta m^2} \cos2\theta}\over{2V T}}.
\label{eres}
\end{equation}
The dependence of resonant neutrino energy on temperature and lepton number is
\begin{eqnarray}
\label{eresapprox}
{\epsilon}_{\rm res}  & \approx & {{{\pi^2}\delta m^2 \cos2\theta}\over{2^{5/2} \zeta\left( 3\right) G_{\rm F} {\left( {\cal{L}}\pm\eta/4\right)}T^4 }}
\\
& \approx & 0.124{\left({{\delta m^2\cos2\theta}\over{1\,{\rm eV}^2}}\right)} {{1}\over{{\cal{L}}}} {{\left( {{\rm MeV}\over{T}}\right)}^4}. \nonumber
\end{eqnarray}
It is clear from Eq.\ (\ref{eresapprox}) that as the universe expands and the
temperature drops, the resonance energy ${\epsilon}_{\rm res}$ will sweep from
lower to higher values. In fact, as the resonance sweeps through the active
neutrino distribution, converting $\nu_\alpha \rightarrow \nu_s$, ${\cal{L}}$
will decrease, further accelerating the resonance sweep rate.

Assuming homogeneity and isotropy, the number density of active neutrinos
$\nu_{\alpha}$ with thermal distribution function
$f_{\nu_\alpha}\left(\epsilon\right)$ in the scaled energy range $\epsilon$ to
$\epsilon +d\epsilon$ is
\begin{equation}
dn_{\nu_\alpha} = n_{\nu_\alpha} f_{\nu_\alpha}\left(\epsilon \right)
d\epsilon ,
\label{no}
\end{equation}
where $n_{\nu_\alpha}$ is the total number density (that is, integrated over
all neutrino energies).  In terms of the temperature $T$ and degeneracy
parameter $\eta_{\nu_\alpha}\equiv {\mu_{\nu_\alpha}}/T$, where
$\mu_{\nu_\alpha}$ is the appropriate chemical potential, the thermal
distribution function is
\begin{equation}
f_{\nu_\alpha}\left(\epsilon \right) =
{{1}\over{F_2\left(\eta_{\nu_\alpha}\right)}}
{{{\epsilon}^2}\over{e^{\epsilon-\eta_{\nu_\alpha}}+1}} .
\label{distrib}
\end{equation}
We define relativistic Fermi integrals of order $k$ in the usual fashion:
\begin{equation}
F_k\left(\eta \right) \equiv \int_0^\infty{ {{x^k dx}\over{e^{x-\eta}+1}} } .
\label{Fermi}
\end{equation}
The total number density of thermally distributed active neutrinos $\nu_\alpha$
with temperature $T_{\nu}$ and degeneracy parameter $\eta_{\nu_\alpha}$ is
\begin{equation}
n_{\nu_\alpha}= {{T_{\nu}^3}\over{2\pi^2}}
F_2\left(\eta_{\nu_\alpha}\right) .
\label{totno}
\end{equation}
Note that if the neutrino degeneracy parameter is $\eta_{\nu_\alpha}=0$, then
$F_2\left( 0\right) =3\zeta\left( 3\right)/2$ and the number density of
thermally distributed $\nu_\alpha$'s is
\begin{equation}
n_{\nu_\alpha} = {{3}\over{8}} n_{\gamma} {\left(
{{T_{\nu}}\over{T_{\gamma}}} \right)}^3,
\label{nnugam}
\end{equation}
 where we allow for the neutrino temperature $T_{\nu}$ to differ from the
 photon/plasma temperature $T_\gamma$.

The relationship between the lepton number in $\alpha$ flavor neutrinos and the
$\nu_\alpha$ degeneracy parameter is
\begin{equation}
L_{\nu_\alpha} = {\left( {{\pi^2}\over{12\zeta\left( 3\right)}} \right)}
{\left( {{T_{\nu}}\over{T_{\gamma}}} \right)}^3 \left[ \eta_{\nu_\alpha}+
\eta_{\nu_\alpha}^3/\pi^2 \right].
\label{leta}
\end{equation}
This relation assumes that neutrinos have Fermi-Dirac energy spectra
and that $\nu_\alpha$ and antineutrinos $\bar\nu_\alpha$ are (or were
at one point) in thermal and chemical equilibrium so that
$\eta_{\bar\nu_\alpha}=-\eta_{\nu_\alpha}$.  In the limit where the
lepton number is small, so that $\eta_{\nu_\alpha} \ll 1$, and the
neutrino and photon temperatures are nearly the same, we can
approximate Eq.\ (\ref{leta}) as $\eta_{\nu_\alpha} \approx 1.46
L_{\nu_\alpha}$. Neutrino degeracy parameter is a comoving invariant;
whereas, lepton number is not in general since the photons can be
heated relative to the neutrinos by, {\it e.g.}, $e^\pm$ annihilation.
 
\subsection{Lepton Number Depletion and 
the Time/Temperature Dependence of Resonance Energies}

As the universe expands and $\nu_\alpha$ neutrinos are converted to sterile
species $\nu_s$, the lepton number $L_{\nu_\alpha}$ drops. As ${\cal{L}}$
approaches zero, the resonance sweep rate becomes so large that neutrinos will
be propagating through MSW resonances non-adiabatically \cite{SF97}. Efficient
neutrino flavor conversion ceases at this point. If the conversion process
results in a change in the number density of $\nu_\alpha$ neutrinos, $\Delta
n_{\nu_\alpha}$, such that the lepton number associated with this species
changes by $\Delta L_{\nu_\alpha}=-\Delta n_{\nu_\alpha}/n_\gamma$, then the
potential lepton number would change from its initial value, ${\cal{L}}^{\rm
initial}$ to
\begin{equation}
{\cal{L}}^{\rm final}={\cal{L}}^{\rm initial}+2\Delta L_{\nu_\alpha}.
\label{Lf}
\end{equation}
The adiabaticity condition ensures that flavor conversion ceases when
${\cal{L}}^{\rm final}$ approaches zero, or in other words, when $\Delta
L_{\nu_\alpha}=-{\cal{L}}^{\rm initial}/2$. It is important to note that
transformation of {\it any flavor} active neutrino to sterile flavor can drive
down the overall potential lepton number, no matter which flavor or flavors of
active neutrinos harbor the net lepton number.

If additionally we were to assume that the resonance smoothly and
continuously swept through the $\nu_\alpha$ energy distribution from
zero to scaled energy $\epsilon$ during this conversion process, we
would have $\Delta n_{\nu_\alpha} =\int_0^{\epsilon}{dn_{\nu_\alpha}}$
and so the concomitant change in lepton number would be
\begin{equation}
\Delta L_{\nu_\alpha} \approx -{{3}\over{8}} {\left( {{T_{\nu}}\over{T_{\gamma}}} \right)}^3 {{1}\over{F_2\left( 0\right) }} \int_0^{\epsilon}{{{x^2}\over{e^{x-\eta_{\nu_\alpha}}+1}}dx  }.
\label{delL}
\end{equation} 
In this idealized scenario, the potential lepton number as a function of
$\epsilon$ is
\begin{equation}
{\cal{L}}{\left(\epsilon\right)} \approx {\cal{L}}^{\rm initial}-{{3}\over{4}} {{1}\over{F_2\left( 0\right) }} \int_0^{\epsilon}{{{x^2}\over{e^{x-\eta_{\nu_\alpha}}+1}}dx  }.
\label{Lofeps}
\end{equation} 
In this last relation we have set the photon/plasma and neutrino temperatures
to be the same. This is a good approximation in the epoch where it turns out we
will be most interested in resonance sweep, between Weak Decoupling and $T
\approx 0.5\,{\rm MeV}$. During this time there has been little annihilation of
$e^\pm$ pairs and, consequently, little heating of the photons/plasma relative
to the decoupled neutrinos.

\begin{figure}
\includegraphics[width=3.25in]{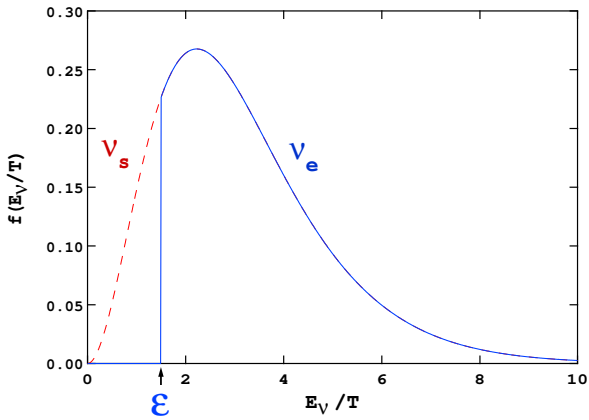}
\caption{The nonthermal scaled energy ($E_\nu/T$) distributions $f\left(
  E_\nu/T\right)$ for $\nu_s$ (dashed) and $\nu_e$ (solid) resulting from
  smooth, adiabatic resonance sweep from $E_\nu/T = 0$ to $E_\nu/T =
  \epsilon$. }
\label{figure1}
\end{figure}

Employing the approximation of a smooth and continuous sweep of scaled
resonance energy from zero to $\epsilon$, we can re-write the resonance
condition, Eq.\ (\ref{eresapprox}), as
\begin{equation}
{\epsilon}{\cal{L}}{\left(\epsilon\right)} \approx {{{\pi^2}\delta m^2
\cos2\theta}\over{2^{5/2} \zeta\left( 3\right) G_{\rm F}T^4 }}.
\label{ereseqn}
\end{equation}

For smooth, continuous and {\it adiabatic} ({\it i.e.}, complete conversion)
resonance sweep up to a scaled energy $\epsilon$, the resulting active and
sterile neutrino distribution functions would be as shown in
Fig.~(\ref{figure1}).  This energy distribution function is zero for all values
of scaled neutrino energy $0\le E_{\nu}/T\le \epsilon$, and has a normal
Fermi-Dirac thermal distribution character for all neutrino energies
$E_{\nu}/T>\epsilon$. The corresponding sterile neutrino energy spectrum would
be the \lq\lq mirror image\rq\rq\ of this: a thermal Fermi-Dirac spectrum for
$0\le E_{\nu}/T\le \epsilon$, and zero occupation for $E_\nu/T >\epsilon$.

However, Eq.\ (\ref{ereseqn}) reveals a problem: the resonance cannot sweep
continuously and smoothly to the point where ${\cal{L}}{\left(\epsilon\right)}
\rightarrow 0$. This is because ${\epsilon}{\cal{L}}{\left(\epsilon\right)}$ is
a peaked function. The maximum of this function occurs for a value
$\epsilon_{\rm max}$ satisfying the integral equation
\begin{equation}
\epsilon_{\rm max}^3\approx 2\zeta\left( 3\right) {\left( e^{\epsilon_{\rm max}-\eta_{\nu_\alpha}}+1 \right)} {\cal{L}}{\left(\epsilon_{\rm max}\right)}. 
\label{emax}
\end{equation}
It is clear, however, that as the universe expands, the right hand side of
Eq.\ (\ref{ereseqn}) will increase monotonically. Although the resonance sweep
can begin smoothly and continuously, there will come a point where it is no
longer possible to find a solution to Eq.\ (\ref{ereseqn}). This will occur
when the resonance energy reaches $\epsilon_{\rm max}$.

\begin{figure}
\includegraphics[width=3.5in]{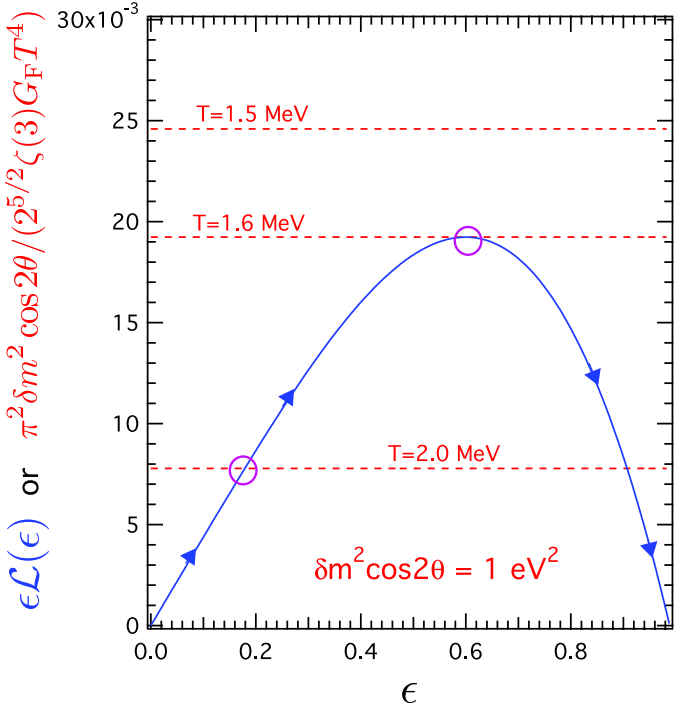}
\caption{The solid line is $\epsilon {\cal{L}}(\epsilon)$, the product of $\epsilon$
  and potential lepton number, in the smooth and continuous resonance sweep
  case for initial lepton numbers $L_{\nu_e}=L_{\nu_\mu}=L_{\nu_\tau}=0.01096$,
  corresponding to initial potential lepton number ${\cal{L}}(\epsilon=0) =
  0.04384$. The arrows give the sense of evolution along this curve as the
  universe expands and the net potential lepton number decreases as a result of
  neutrino flavor conversion in the channel $\nu_e\rightarrow \nu_s$ with
  $\delta m^2\cos2\theta=1\,{\rm eV}^2$.  The horizontal dashed lines
  correspond to values of the right-hand side of Eq.~(\ref{ereseqn}) for the
  indicated epochs (temperatures).  Solutions to Eq.~(\ref{ereseqn}) are
  possible at a given epoch when the corresponding dashed line crosses the
  $\epsilon {\cal{L}}(\epsilon)$ curve.  Physical solutions are circled here for
  $T=2.0\,{\rm MeV}$ and $1.6\,{\rm MeV}$.  The maximum value on the $\epsilon
  {\cal{L}}(\epsilon)$ curve occurs at $\epsilon_{\rm max}\approx 0.598$, and in the
  smooth resonance sweep scenario this is reached at $T\approx 1.6\,{\rm
    MeV}$. Clearly, no solutions are possible in this scenario for $T <
  1.6\,{\rm MeV}$. If the system were forced to follow the smooth resonance
  sweep $\epsilon {\cal{L}}(\epsilon)$ curve beyond $\epsilon_{\rm max}$, the potential
  lepton number would be completely depleted when $\epsilon$ reaches
  $\epsilon_{\rm c.o.}\approx 0.987$ ({\it i.e.}, ${\cal{L}}(\epsilon_{\rm
    c.o.})=0)$. }
\label{figure2b}
\end{figure}

Fig.~(\ref{figure2b}) shows graphically the problem of obtaining a solution to
Eq.~(\ref{ereseqn}) for a particular case. The solid line in this figure is
$\epsilon {\cal{L}}(\epsilon)$ as computed by assuming a smooth and continuous
resonance sweep scenario.  Here we have chosen initial lepton numbers
$L_{\nu_e}=L_{\nu_\mu}=L_{\nu_\tau}=0.01096$. This corresponds to an initial
potential lepton number ${\cal{L}}(\epsilon=0) = 0.04384$. The arrows in this
figure give the sense of evolution along the solid curve as the universe
expands and the net potential lepton number decreases as a result of neutrino
flavor conversion in the channel $\nu_e\rightarrow \nu_s$. The maximum value on
this curve occurs at $\epsilon_{\rm max}\approx 0.598$.

However, the potential lepton number would be completely depleted ({\it i.e.},
${\cal{L}}(\epsilon_{\rm c.o.})=0)$ in the smooth and continuous resonance
sweep scenario only when $\epsilon$ reaches a \lq\lq cut-off\rq\rq\ value
$\epsilon_{\rm c.o.}$, which is $\approx 0.987$ in this case.  The horizontal
dashed lines in Fig.~(\ref{figure2b}) correspond to values of the right-hand
side of Eq.\ (\ref{ereseqn}) for epochs of the universe corresponding to
temperatures $T=2.0\,{\rm MeV}$, $T=1.595\,{\rm MeV}$, and $T=1.5\,{\rm MeV}$.
Solutions to this equation are possible when these curves cross the solid
$\epsilon L(\epsilon)$ curve.  Clearly, no solutions are possible in the
continuous resonance sweep scenario for $T < 1.6\,{\rm MeV}$ in this case.

What happens beyond this point, {\it e.g.,} for $T < 1.6\,{\rm MeV}$?  If we
relax the demand that the resonance sweep be continuous, then it is possible in
principle to find a solution to Eq.\ (\ref{ereseqn}) as the temperature drops
beyond the point where $\epsilon=\epsilon_{\rm max}$, though this would require
that the product of scaled resonance energy and potential lepton number differ
from the solid curve $\epsilon {\cal{L}}{\left(\epsilon\right)}$.  Though a
detailed numerical model of this process is beyond the scope of this work, we
can get a rough idea of what might happen with the following argument.

Suppose we take a time step resulting in a new temperature $T^\prime$ slightly
lower than $T_{\rm max}$, the temperature where the last continuous sweep
solution exists, {\it i.e.,} where $\epsilon=\epsilon_{\rm max}$ ($T_{\rm
  max}\approx 1.6\,{\rm MeV}$ for our example case) and
${\cal{L}}={\cal{L}}(\epsilon_{\rm max})$.  One possibility is that the
resonance energy could skip to some value $\epsilon>\epsilon_{\rm max}$, toward
the higher energy portion of the neutrino distribution function.  In this way
the product $\epsilon {\cal{L}}(\epsilon_{\rm max})$ could be large enough to
match the right-hand side of Eq.\ (\ref{ereseqn}) at the new temperature
$T^\prime$.  Of course, this would result in resonant neutrino flavor
conversion and so ${\cal{L}}$ would be lowered and eventually we again would be
unable to maintain a smooth resonance sweep. At that point the resonance energy
could skip again discontinuously, {\it etc.}.  It is possible that beyond
$\epsilon_{\rm max}$ the resonance sweeps stochastically in this way through
relatively small intervals of the active neutrino distribution function leaving
a \lq\lq picket fence\rq\rq\ distribution beyond $\epsilon_{\rm max}$.

Though the details may differ from this simple scheme, we believe that resonant
neutrino flavor conversion for $\epsilon>\epsilon_{\rm max}$ {\it will} occur
because: (1) as we show below, neutrino flavor evolution at this epoch is very
adiabatic for relevant neutrino parameters; and (2) the resonance condition can
be met for some value of neutrino energy so long as the net lepton number is
non-zero.  In any case, however, active-to-sterile neutrino conversion
$\nu_\alpha\rightarrow\nu_s$ will have to cease when ${\cal{L}}$ approaches
zero.

At this point we will be left with grossly non-thermal, non-Fermi-Dirac
$\nu_\alpha$ and $\nu_s$ distributions. Since this process occurs after Weak
Decoupling, active neutrino inelastic scattering processes on electrons,
nucleons, and other neutrinos have rates which are slow compared to the
expansion rate of the universe. This means that these processes will be unable
to redistribute effectively the active neutrino occupation numbers and so they
cannot morph the $\nu_\alpha$ distribution into a thermal distribution. This
has consequences for the lepton capture rates on nucleons as we will discuss
below in section IV.

In either the (unphysical) smooth and continuous resonance sweep scenario or in
some stochastic resonance sweep case both the active neutrino and resulting
sterile neutrino distribution functions will be non-thermal in character.
Since we cannot solve for resonance sweep beyond $\epsilon_{\rm max}$ we do not
know the final active and sterile neutrino energy spectra. However, for the
purposes of constraints and general guidelines, we will find that the idealized
smooth and continuous resonance sweep scenario provides the basis for lower
limits on the effects of neutrino spectral distortion.
 
\begin{figure}
\includegraphics[width=3.25in]{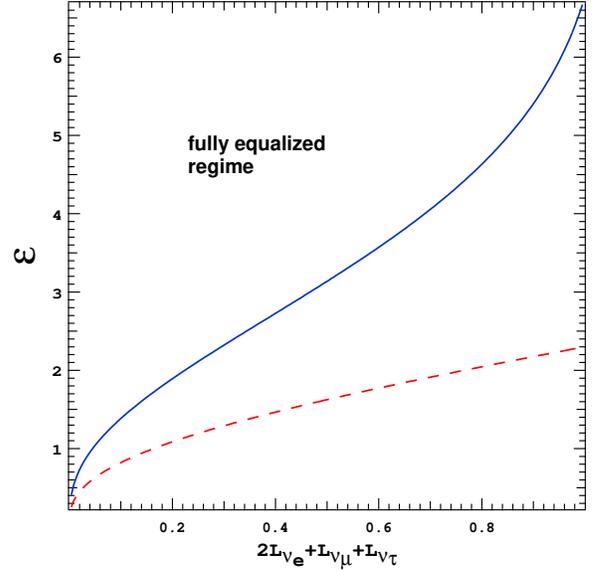}
\caption{The values of $\epsilon_{\rm c.o.}$ (solid line) and $\epsilon_{\rm
    max}$ (dashed line) are shown as functions of total initial potential
  lepton number in the limit of a smooth and continuous resonance sweep and
  with the assumption that full active neutrino equilibration obtains
  ($L_{\nu_e}=L_{\nu_\mu}=L_{\nu_\tau}$).  }
\label{figure2}
\end{figure}

It is useful to consider the solution for the cut-off energy $\epsilon_{\rm
  c.o.}$ and the peak energy $\epsilon_{\rm max}$ in the smooth and continuous
resonance sweep case. To get the first of these quantities, we force the system
to evolve continuously ({\it e.g.}, along the solid curve in
Fig.~\ref{figure2b}) all the way to complete lepton number depletion and solve
\begin{equation}
{\cal{L}}^{\rm initial} ={{3}\over{4}} {{1}\over{F_2\left( 0\right) }} \int_0^{\epsilon_{\rm c.o.}}{{{x^2}\over{e^{x-\eta_{\nu_\alpha}}+1}}dx  }, 
\label{sol}
\end{equation}
or ${\cal{L}}\left(\epsilon_{\rm c.o.}\right)=0$. The second of these quantities is 
the solution of Eq.\ (\ref{emax}). Both of these solutions are shown
as functions of initial potential lepton number, ${\cal{L}}^{\rm initial}$
in Fig.\ (\ref{figure2}). In this figure
it is assumed that the active neutrinos are fully
\lq\lq equilibrated\rq\rq\ initially (before any flavor transformation) with
$L_{\nu_e}=L_{\nu_\mu}=L_{\nu_\tau}$.

It is obvious from Eq.\ (\ref{sol}) that there is a maximum value of the
initial potential lepton number for which a solution is obtainable when
$\nu_\alpha \rightleftharpoons \nu_s$ is the only operative neutrino flavor
conversion channel. This maximum is given by the limit where $\epsilon_{\rm
  c.o.}\rightarrow \infty$,
\begin{equation}
{\cal{L}}^{\rm initial}_{\rm max}\approx
 {{3}\over{4}}{{F_2\left(\eta_{\nu_\alpha}\right)}\over{F_2\left(0\right) }}.
\label{asymp}
\end{equation}
Scenarios where the bulk of the initial potential lepton number is contained in
seas of another flavor of active neutrinos may not allow
$\nu_\alpha\rightleftharpoons\nu_s$ conversion to leave a zero final potential
lepton number. This is a simple consequence of the post Weak Decoupling
conservation of numbers of neutrinos of all kinds. Of course, active-active
neutrino flavor transformation in the channels
$\nu_\alpha\rightleftharpoons\nu_\beta$ ($\alpha,\beta={\rm e},\mu,\tau$) can
alter this picture significantly and will be discussed below.

\subsection{Efficiency of Neutrino Flavor Conversion: Adiabaticity}

From the previous discussion it is clear that the efficiency of active-sterile
neutrino flavor conversion at MSW resonances is a key issue in resolving how
these resonances sweep with scaled energy $\epsilon$.  This is especially true
for values of scaled resonance energy beyond $\epsilon_{\rm max}$, where we
argued that {\it if} the resonance condition can be met neutrino flavor
transformation was likely to be efficient for the typical neutrino mass/mixing
parameters we consider here. It is the adiabaticity of neutrino propagation
which determines transformation efficiency both in the active-sterile and
active-active channels.  We point out here that adiabaticity parameters are
high for our chosen epoch and neutrino mass/mixing characteristics, essentially
because these parameters are proportional to the ratio of a gravitational time
scale (the causal horizon) to a weak timescale (in-medium oscillation time).

The causal horizon (particle horizon) is the proper distance traversed by a
null signal in the age of the universe $t$. In radiation-dominated conditions
in the early universe this is (setting $c=1$)
\begin{equation}
d_{\rm H}\left(t\right) = 2t=H^{-1},
\label{horizon}
\end{equation} 
where the local Hubble expansion rate is
\begin{equation}
H\approx {\left( {{8\pi^3}\over{90}} \right)}^{1/2} g^{1/2} {{T^2}\over{m_{\rm
      pl}}}.
\label{hubble}
\end{equation} 
Here $m_{\rm pl} \approx 1.221\times{10}^{22}\,{\rm MeV}$ is the Planck mass.
The statistical weight for a relativistic boson species $i$ is $\left(
g_b\right)_i$, while that for a relativistic fermion species $j$ is
$\left(g_f\right)_j$.  These are related to the total statistical weight $g$ by
a sum over all particle species $i$ and $j$ with relativistic kinematics with
equilibrium or near equilibrium energy distribution functions and energy
densities in the plasma at temperature $T$, given by $g \equiv
\sum_i{\left(g_b\right)}_i+(7/8)\sum_j{\left(g_f\right)}_j$. In the epoch
between Weak Decoupling and Weak Freeze Out and BBN, photons, $e^\pm$ pairs and
the active neutrinos are relativistic and appreciably populated so that
$g\approx 10.75$ and $t\approx \left( 0.74\,{\rm s}\right)
{\left(10.75/g\right)}^{1/2} {\left({\rm MeV}/T\right)}^2$.  The spectral
distortions and extra energy density stemming from the net lepton numbers
considered in this paper are usually small effects, causing deviations of the
expansion rate from that given above by less than a few percent in many cases.

Homogeneity and isotropy in the early universe imply that the entropy in a
co-moving volume is conserved. The proper, physical entropy density in
radiation-dominated conditions is $S\approx {\left( 2\pi^2/45\right)} g_s T^3$,
where $g_s$ is closely related to $g$ and we can take $g_s\approx g$. We can
take the co-moving volume element to be the cube of the scale factor $a$ in the
Friedman-Lemaitre-Robertson-Walker (FLRW) metric, so that $a^3 S$ is invariant
with FLRW time coordinate $t$ and therefore $g^{1/3} a T$ is constant. In turn,
this implies that the fractional rate of change of the temperature is related
to the expansion rate and the fractional rate of change of the statistical
weight by
\begin{equation}
{{\dot{T}}\over{T}} \approx -H{\left( 1+{{{\dot{g}}/{g}}\over{3H}}\right)}.
\label{comovings}
\end{equation}

At lower temperatures, where the thermal potential can be neglected, the
potential governing neutrino flavor transformation is the difference of the
Hamiltonians ({\it e.g.},
Eq.s\ (\ref{hnue}),(\ref{hnumu}),(\ref{hnutau}),(\ref{hsterile})) for the
transforming neutrino species. For the active-sterile channel
$\nu_\alpha\rightleftharpoons\nu_s$, for example, we have $V\approx
H(\nu_\alpha)-H(\nu_s)$. The appropriate potentials for the active-active
neutrino flavor transformation channels follow in like manner.

The density scale height for the early universe depends on the neutrino flavor
transformation channel and is defined as
\begin{eqnarray}
\label{densityscale}
{\cal{H}} & \equiv & {\bigg\vert {{1}\over{V}} {{dV}\over{dt}}\bigg\vert}^{-1}
\\ & \approx & {{1}\over{3}} H^{-1} {\bigg\vert 1+{{{\dot{g}}/{g}}\over{3H}}
  -{{{\dot{{\cal{L}}}}/{{\cal{L}}}}\over{3H}}
  \bigg\vert}^{-1}.\nonumber
\end{eqnarray} 
Here the approximation on the
second line is for active-sterile neutrino flavor transformation and follows on
neglecting the thermal potential $C$ and using Eq.\ (\ref{comovings}). When the
statistical weight and the lepton numbers are not changing rapidly the density
scale height is roughly a third of the horizon scale. This is $\sim
{10}^5\,{\rm km}$ at the epoch we are considering here.

Define $\Delta\equiv \delta m^2/2E_\nu$. It can be shown that the ratio of the
difference of the squares of the {\it effective} masses in matter to twice the
neutrino energy is
\begin{eqnarray}
\label{DeltaEff}
\Delta_{\rm eff} & \equiv & {{\delta m^2_{\rm eff}}\over{2 E_\nu}} \\ & \approx
& \sqrt{{\left(\Delta\cos2\theta-V \right)}^2+{\left( \Delta
    \sin2\theta+B_{e\tau} \right)}^2 },\nonumber
\end{eqnarray}
where $\theta$ is the appropriate effective two-by-two vacuum mixing angle and
where $V=A+B+C$ is the appropriate potential for the transformation
channel. Here $B_{\rm e\tau}$ is the flavor-off diagonal potential as defined
in Qian \& Fuller 1995 \cite{QF95}. The flavor basis off-diagonal potential
vanishes, $B_{e\tau}=0$, for any active-sterile mixing channel.

The effective matter (in-medium) mixing angle $\theta_{\rm M}$ for a neutrino
transformation channel with potential $V$ and effective vacuum mixing angle
$\theta$ satisfies
\begin{widetext}
\begin{equation}
\sin^22\theta_{\rm M} = {{\Delta^2 \sin^22\theta {\left( 1+2E_\nu B_{e\tau}/\delta m^2 \sin2\theta \right)}^2}\over{{{\left(\Delta\cos2\theta-V  \right)}^2+\Delta^2 \sin^22\theta{\left(1+2E_\nu B_{e\tau}/\delta m^2 \sin2\theta \right)}^2 }  }}.
\label{mattermix}
\end{equation}
\end{widetext}
The effective matter mixing angle for the antineutrinos in this channel,
$\bar\theta_{\rm M}$, satisfies a an expression which has opposite signs for
the potentials $B$, $A$, and $B_{e\tau}$, but which is otherwise identical.

The change in the potential required to drop the effective matter mixing from
the maximal resonant value ($\theta_{\rm M} =\pi/4$) to a value where
$\sin^22\theta_{\rm M} = 1/2$ is termed the resonance width and is
\begin{equation}
\delta V \approx \Delta \sin2\theta {\bigg\vert 1+{{2E_\nu
      B_{e\tau}}\over{\delta m^2 \sin2\theta}}
  \bigg\vert}.
\label{width}
\end{equation} 
The physical width in space, or
in FLRW coordinate time $t$, corresponding to this potential width is
\begin{eqnarray}
\label{timewidth}
\delta t & = & {{dt}\over{dV}} \delta V\approx {\bigg\vert {{1}\over{V}}
  {{dV}\over{dt}}\bigg\vert}^{-1} {{\delta V}\over{V}}\bigg\vert_{\rm res} \\ &
\approx & {\cal{H}} \tan2\theta {\bigg\vert 1+{{2E_\nu B_{e\tau}}\over{\delta
      m^2 \sin2\theta}} \bigg\vert}.\nonumber 
\end{eqnarray}

The local neutrino oscillation length at resonance is
\begin{equation}
L_{\rm osc}^{\rm res} = {{4\pi E_\nu}\over{\delta m^2_{\rm
      eff}}}={{2\pi}\over{\Delta_{\rm eff}}} \approx {{2\pi}\over{\delta V}},
\label{osclength}
\end{equation}
where the latter approximation is good only at resonance. We can define the dimensionless adiabaticity parameter as proportional to the ratio of the resonance width and the neutrino oscillation length at resonance:
\begin{eqnarray}
\label{adparam}
\gamma & \equiv & 2\pi{{\delta t}\over{L_{\rm osc}^{\rm res}}}\approx {\delta
  t}{\delta V} \\ & \approx & {{\delta m^2 {\cal{H}}}\over{2E_\nu}}\cdot
       {{\sin^22\theta}\over{\cos2\theta}}\cdot {\bigg\vert 1+{{2E_\nu
             B_{e\tau}}\over{\delta m^2 \sin2\theta}}
         \bigg\vert}^2.\nonumber\end{eqnarray} 
This parameter can be evaluated anywhere in the evolution of neutrino flavors,
even well away from resonances and it will serve to gauge the degree to which
neutrinos tend to remain in mass eigenstates. The Landau-Zener jump
probability, assuming a linear change in potential across the resonance width,
is $P_{\rm LZ} \approx \exp{\left(-\pi\gamma/2\right)}$, so that it is clear
that a large value of the adiabaticity parameter corresponds to a small
probability of jumping between mass eigenstate tracks and, hence, efficient
flavor conversion at asymptotically large distance (many resonance widths) from
resonance.

Folding in the expansion rate in radiation-dominated conditions, using the
conservation of co-moving entropy density, and assuming that we can neglect the
thermal potential $C$, we can show that the adiabaticity parameter for neutrino
propagation through an active-sterile resonance is
\begin{widetext}
\begin{eqnarray}
\label{adnumb}
\gamma & \approx & {{\sqrt{5}\,{\zeta^{3/4}\left(3\right)}}\over{2^{1/8} \pi^3}} \cdot {{{\left(\delta m^2\right)}^{1/4} m_{\rm pl} G_{\rm F}^{3/4}}\over{g^{1/2}}} \cdot 
{\left[{{\cal{L}}^{3/4}}\over{\epsilon^{1/4}}\right]}
\cdot {\left[{{\sin^22\theta}\over{\cos^{7/4}2\theta}}\right]}\cdot{ {\bigg\vert 1+{{{\dot{g}}/{g}}\over{3H}}
-{{{\dot{{\cal{L}}}}/{{\cal{L}}}}\over{3H}} \bigg\vert} }^{-1}  \\
& \approx & {\left( {{10.75}\over{g}} \right)}^{1/2}\cdot{\left[ {{\delta m^2}\over{1\,{\rm eV}^2}} \right]}^{1/4}\cdot{{1}\over{\epsilon^{1/4}}}\cdot{\left[ {{{\cal{L}}}\over{0.01}} \right]}^{3/4}\cdot{ {\bigg\vert 1+{{{\dot{g}}/{g}}\over{3H}}
-{{{\dot{{\cal{L}}}}/{{\cal{L}}}}\over{3H}} \bigg\vert} }^{-1}
\cdot{\left\{ {{\sin^22\theta}\over{1.77\times{10}^{-8}}} \right\}}.\nonumber
\end{eqnarray}
\end{widetext}
In these expressions $\epsilon=E_\nu/T$ is the scaled energy of a neutrino at
resonance in a channel $\nu_\alpha\rightleftharpoons\nu_s$ characterized by the
difference of the squares of the appropriate vacuum mass eigenvalues, $\delta
m^2$. It is obvious from these considerations that neutrino flavor
transformation will be efficient at resonance ({\it i.e.}, $\gamma \gg 1$) over
broad ranges of energy for the regime of the early universe between Weak
Decoupling and Weak Freeze Out even for very small effective vacuum mixing
angle $\theta$.

Eq.\ (\ref{adnumb}) shows that two trends can eventually destroy adiabaticity
and, therefore, large scale resonant active-sterile neutrino flavor
transformation. As active neutrinos are converted ${\cal{L}}$ is reduced and
this reduces $\gamma$. In turn, the fractional rate of destruction of
${\cal{L}}$ compared with the Hubble parameter can be become significant,
especially if ${\cal{L}}$ is small, and this can also reduce $\gamma$.

\subsection{Active-active Neutrino Flavor Conversion and Equilibration}

Active neutrinos
($\nu_e$,$\bar\nu_e$,$\nu_\mu$,$\bar\nu_\mu$,$\nu_\tau$,$\bar\nu_\tau$)
transforming among themselves on time scales comparable to or shorter than that
of the active-sterile conversion channel can alter significantly the scenario
for sterile neutrino production given above. This is apt to be the case if
active-active neutrino mixing in medium is large and efficient over a broad
range of neutrino energies. Active-sterile neutrino flavor conversion tends to
be slow because it occurs through MSW resonances and the rate at which these
resonances sweep through the neutrino distribution functions is determined by
the expansion of the universe, a slow gravitational time scale.

Coherent neutrino flavor conversion in active-active channels in the early
universe can be dominated by the flavor off-diagonal potential. Large in-medium
mixing angles can accompany the synchronization seen in calculations of
active-active mixing in supernovae and the early universe \cite{abb}. If
active-active neutrino flavor transformation is efficient, then lepton numbers
in different active neutrino species can be quickly equilibrated, meaning
instantaneous equal lepton numbers.

\begin{figure}
\includegraphics[width=3.25in]{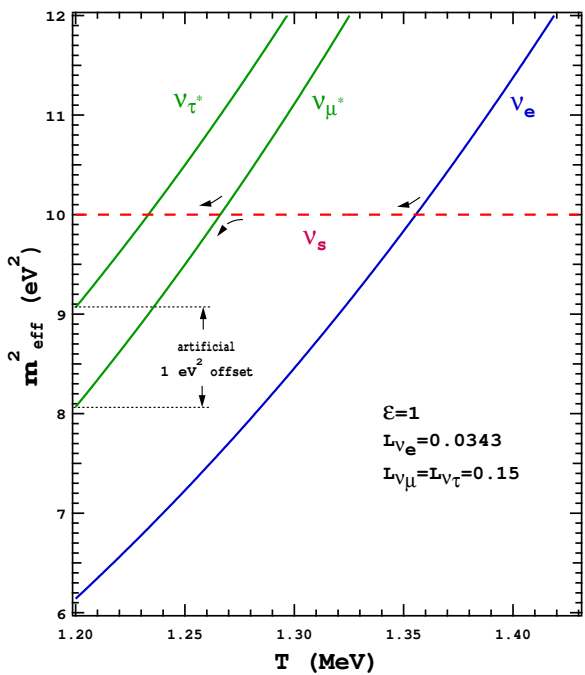}
\caption{Level crossing diagram for the case with lepton numbers as shown and
  for scaled neutrino energy $\epsilon =1$.  The vacuum mass-squared eigenvalue
  for the (mostly) sterile state is taken as $m^2_4 = 10\,{\rm eV}^2$. This is
  shown as the dashed curve labeled $\nu_s$.  An artificial (exaggerated)
  $1\,{\rm eV}^2$ offset between the vacuum mass-squared eigenvalues $m_2^2$
  and $m_3^2$ has been added so that the curves labeled with $\nu_\mu^\ast$ and
  $\nu_\tau^\ast$ are separated for clarity. In reality, the top curve should
  be split from the lower curve by $\delta m^2\approx 3\times{10}^{-3}\,{\rm
    eV}^2$.  Conversion in the channel $\nu_e\rightarrow\nu_s$ is as described
  in the text. }
\label{figure3}
\end{figure}
The flavor diagonal neutrino forward scattering potential in an active-active
channel $\nu_{\alpha}\rightleftharpoons\nu_\beta$ is
$A+B=H\left(\nu_\alpha\right)-H\left(\nu_\beta\right)$.  If there is an initial
disparity in lepton number in these two flavors then matter-enhanced or
-suppressed transformation will go in the direction of reducing this disparity.
Though initially the flavor off-diagonal potential $B_{e\tau}\approx 0$, as
soon as flavor transformation begins this potential comes up.

The interplay of matter-enhanced coupled active-sterile and active-active
neutrino flavor transformation can be complicated and difficult to follow
numerically. The size of the debit in the $\nu_e$ or $\bar\nu_e$ distributions
({\it e.g.}, the final value of $\epsilon_{\rm c.o.}$ or $\epsilon_{\rm max}$
in a continuous sweep scenario) may be much more complicated in the general
$4\times 4$ case than the scenario outlined above for \lq\lq
simple\rq\rq\ $2\times 2$ $\nu_\alpha\rightleftharpoons \nu_s$
interconversion. We can, however, identify a few cases where we can at least
outline the course of neutrino flavor conversion as the universe expands and
cools.  We will therefore consider two limits: (1) no active-active mixing; and
(2) efficient active-active mixing that guarantees that lepton numbers in
active species are always the same (instantaneous equilibration).

\subsection{Inefficient Active-Active Neutrino Flavor Conversion} 

Consider first the case where we neglect active-active neutrino mixing
effects. In this case we could have initial lepton numbers that are not fully
equilibrated. For example, we could have a scenario where initially $L_{\nu_e}
< L_{\nu_\mu}= L_{\nu_\tau}$.  In this case the $\nu_\mu$ and $\nu_\tau$
experience the largest effective potentials, and hence have the largest
effective masses at a given temperature (epoch) in the early universe.
Therefore, the first (highest temperature) resonance occurs for $\nu_s$ with
$\nu_e$, as illustrated in Fig.~(\ref{figure3}). This resonance will destroy
lepton number, as will the subsequent $\nu_\mu^\ast\rightleftharpoons\nu_s$
resonance, and it will leave a distorted $\nu_e$ spectrum.

Here we follow Ref. \cite{BF} and define linear combinations of the muon and tauon
neutrino flavor states
\begin{eqnarray}
\label{mustar}
\vert \nu_\mu^\ast\rangle & \equiv & {{\vert
    \nu_\mu\rangle-\vert\nu_\tau\rangle}\over{\sqrt{2}}} \\
\vert
\nu_\tau^\ast\rangle & \equiv & {{\vert
    \nu_\mu\rangle+\vert\nu_\tau\rangle}\over{\sqrt{2}}}.
\end{eqnarray} 
This reduces the $4\times 4$ mixing problem of three active neutrinos and a
sterile neutrino into a $3\times 3$ problem with $\vert \nu_\tau^\ast\rangle$
decoupled (a mass eigenstate in vacuum with no mixing with the other
neutrinos). This reduction in dimensionality of the neutrino mixing problem
works in vacuum only if the muon and tauon neutrinos are maximally mixed. It
will be valid in medium only if, additionally, these two neutrino flavors
experience identical matter interactions. This latter condition is met if
$L_{\nu_\mu}=L_{\nu_\tau}$. This symmetry condition will be respected so long
as muon and tauon neutrinos behave and transform identically. Indeed, the
second resonance encountered as the universe cools,
$\nu_\mu^\ast\rightleftharpoons\nu_s$, respects this condition as the
$\vert\nu_\mu^\ast\rangle$ state consists of equal parts muon and tauon
states.

The sterile neutrinos produced through the $\nu_e \rightarrow \nu_s$ resonance
are subsequently transformed into $\nu_\mu^*$ at the second resonance at lower
temperature, as depicted for a particular set of initial lepton numbers in
Fig.~(\ref{figure3}).  This resonance also converts the $\nu_\mu^*$ into the
sterile state, so that the final abundance of sterile neutrinos results from
the conversion of neutrinos which were originally in the $\nu_\mu$ and
$\nu_\tau$ distributions. Since these distributions have higher lepton number
than resides in the $\nu_e$/$\bar\nu_e$ seas, the final number density of
sterile neutrinos will be larger than the number of $\nu_e$ missing from the
$\nu_e$-distribution. As we will see, this case may be more likely to be in
conflict with massive neutrino dark matter constraints.

If we temporarily ignore the effect of active-active neutrino flavor
transformations, then we can make some general statements about the change in
the lepton numbers and $\epsilon$ for the $\nu_e$ or $\bar\nu_e$ distributions
in this case of unequal $L_{\nu_e}$ and $L_{\nu_\mu}=L_{\nu_\tau}$. If the
potential lepton number for electron flavor neutrinos ${\cal{L}}_e$ is driven
to zero first then the changes in the individual active neutrino lepton numbers
must be related by
\begin{equation}
2{\Delta}L_{\nu_e}+{\Delta}L_{\nu_\mu^\ast}=-{\cal{L}}_{e}^{\rm
  initial},
\label{uneq}
\end{equation}
where here $L_{\nu_\mu^\ast}=\left(L_{\nu_\mu}+L_{\nu_\tau}\right)/2$, and
$\Delta L_{\nu_e}=\Delta n_{\nu_e}/n_\gamma$ and $\Delta L_{\nu_\mu} =\Delta
n_{\nu_{\mu}}/n_\gamma$.

Of course, if ${\cal{L}}_e$ is driven to zero before ${\cal{L}}_\mu^\ast$, then
conversion of $\nu_e$'s at the first resonance will cease while conversion in
the channel $\nu_\mu^\ast\rightarrow\nu_s$ continues until
${\cal{L}}_{\mu^\ast}$ is reduced to zero. This will leave ${\cal{L}}_e <0$
which will result in anti-electron neutrino transformation
$\bar\nu_e\rightarrow\bar\nu_s$, leaving a non-thermal deficit in the
$\bar\nu_e$ distribution.

This is likely temporary, however. The $\nu_e$ potential is zero at
the point where ${\cal{L}}_e$ first vanishes. Thereafter, with
reduction in ${\cal{L}}_{\mu^\ast}$, the $\nu_e$ potential's {\it
  magnitude} first increases, but then decreases as the universe
expands and the temperature drops. This can be seen from
Eq.~\ref{eres} and on noting that the potential behaves like $V \sim
{\cal{L}}_e T^3$. The evolution of the potential with time is
determined by the competition between two effects. The conversion of
lepton number discussed above makes ${\cal{L}}_e$ more negative and
larger in magnitude, while the expansion of the universe decreases
$T$. Therefore, the antineutrinos could experience two resonances: (1)
first when ${\cal{L}}_e$ becomes sufficiently negative that the
potential $V$ becomes large enough in magnitude to satisfy
Eq.~(\ref{eres}); and (2) subsequently when the temperature drops
enough that this condition is again satisfied. At the first resonance
we have $\bar\nu_e\rightarrow\bar\nu_s$, but these steriles are
reconverted at the second resonance,
$\bar\nu_s\rightarrow\bar\nu_e$. This can also be viewed from the
perspective of resonance sweep. Note that $\epsilon_{\rm res}$ is
infinite when ${\cal{L}}_e$ first crosses zero, but then decreases as
${\cal{L}}_e$ becomes larger in magnitude as lepton number is
converted, but then \lq\lq turns around\rq\rq\ and begins to sweep
toward higher energy again once $T$ becomes low enough. This process
is directly analogous to the re-conversion of $\bar\nu_s$ neutrinos in
neutrino-heated outflow in supernovae \cite{MFBF}.

In this scenario it is the mu and tau neutrinos, ultimately, that are converted
to sterile neutrinos so that the numbers and kinds of converted active
neutrinos are given by
\begin{equation}
\label{Lnoaa}
{\cal{L}}_{\mu^\ast}^{\rm init} \approx {{2}\over{n_\gamma}} {\left( \Delta
  n_{\nu_\mu^\ast}+\Delta n^\prime_{\nu_{\mu}^\ast}
  \right)}-{{1}\over{n_\gamma}} {\left(\Delta n_{\nu_e}+\Delta n_{\bar\nu_e}
  \right)},
\end{equation} where $\Delta n_{\nu_\mu^\ast}$ and $\Delta
n^\prime_{\nu_{\mu}^\ast}$ are the number of $\nu_{\mu}^\ast$ neutrinos
converted before and after ${\cal{L}}_e$ first vanishes,
respectively. Likewise, $\Delta n_{\nu_e}$ electron neutrinos are converted
before ${\cal{L}}_e$ first vanishes and $\Delta n_{\bar\nu_e} $ electron
antineutrinos afterward, though these $\bar\nu_e$'s are eventually returned to
the distribution.

There is an additional complication: in the case that all three lepton numbers
are equal, the mass-squared differences between the active states are
approximately given by their vacuum values, which are quite small.  The three
resonances depicted in Fig.~(\ref{figure3}) will then be very close together,
and in fact may overlap if the resonance width is sizable.

If the resonances do not overlap, the lepton number destroying resonance will
take place between $\nu_1$ and $\nu_s$, where $\nu_1$ is the lightest neutrino
mass eigenstate.  Since $\nu_1$ has a large $\nu_e$ component, this will leave
a non-thermal $\nu_e$ distribution, and in addition there will be smaller
non-thermal distortions of the $\nu_\mu$ and $\nu_\tau$ spectra.  In the case
that the resonances do overlap, the full details of the evolution will be quite
complicated, but a similar outcome is obtained nonetheless.  To summarize, in
all cases a non thermal $\nu_e$ spectrum results.

\subsection{Efficient Active-Active Mixing: Instantaneously Equilibrated Lepton Numbers} 

Let us now consider the limit where in addition to the active-sterile MSW
transitions, oscillations/transformations between/among the three active
neutrinos occur simultaneously and are efficient. If active-active mixing among
all the active flavors is instantaneous and efficient then we only need to
consider the case where the lepton numbers are equal, $L_e = L_\mu = L_\tau$,
both initially and as active-sterile transformation proceeds.  It has been
shown that large angle mixing between the three active neutrino species results
in the system being driven toward such an equilibrated state \cite{abb} at a
temperature of $T \agt 2$ MeV.
 
An obvious additional effect of efficient active-active oscillations will be to
partially refill any hole that was left in the $\nu_e$ distribution.  It is
important to note, though, that this refilling cannot be complete.  For maximal
$\nu_e-\nu_{\mu,\tau}$ mixing, the hole in the distribution can be only
partially refilled.  In vacuum the measured solar neutrino mixing angle is less
than maximal, $\theta_{\rm solar} \simeq 32.5^\circ$, and $U_{e3}$ is
relatively small. In medium, at best we will obtain maximal matter mixing
angles in the limit where the flavor off-diagonal potential is large. We again
therefore expect about $0$ to $2/3$ refilling at most, so that a non-thermal
$\nu_e$ spectrum is always obtained by the epoch of Weak Freeze Out.  Even if
it were somehow possible for the resonance to effectively involve only $\nu_s$
and $\nu_\mu$/$\nu_\tau$, active-active oscillations would again act to refill
the hole in the resulting non thermal $\nu_\mu$ and/or $\nu_\tau$ spectra, and
in so doing create a non-thermal $\nu_e$ distribution.

\begin{figure}
\includegraphics[width=3.25in]{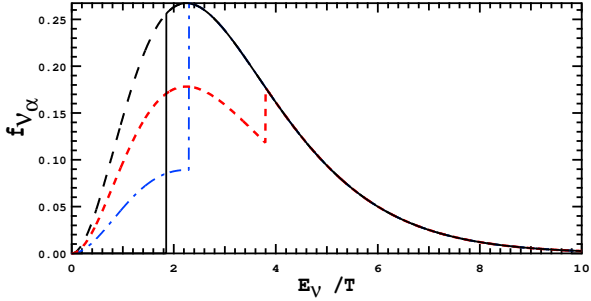}
\caption{Final active neutrino energy distribution function $f_{\nu_\alpha}$
  for Cases 1 (short dashed line), 2 (dot dashed line), and 3 (solid line) for
  the particular case of a continuous resonance sweep scenario and in the
  instantaneous active-active mixing limit as described in the text.  Here
  $\alpha = {\rm e},\mu,\tau$: all species have the same distribution function.
  The long dashed line and its continuation as a solid line shows the original
  thermal distribution function common to all active flavors. The particular
  scenario shown here has $L_{\nu_e}=L_{\nu_\tau}=L_{\nu_\mu}= 0.1$, so that
  $\epsilon_1 \approx 3.8$, $\epsilon_2 \approx 2.3$, and $\epsilon_3\approx
  1.85$ }
\label{figure4}
\end{figure}

We can identify three cases. 

Case 1: We have only one sterile neutrino species and only one channel for its
production, $\nu_\alpha\rightarrow \nu_s$. If this channel is, {\it e.g.},
$\nu_e \rightarrow \nu_s$, then the neutrinos in the $\nu_\tau$ and $\nu_\mu$
distributions will, in the limit of instantaneous maximal mixing, partially
fill in the hole left by the active-sterile conversion process. Given the
boundary condition of equal lepton numbers in all active flavors at all times,
a continuous smooth resonance sweep scenario will leave each active neutrino
distribution with a low energy \lq\lq hole\rq\rq\ with $2/3$ of the normal
population out to some value of scaled neutrino energy $\epsilon_1$. In terms
of the initial potential lepton number ${\cal{L}}^{\rm
  init}={\cal{L}}_e={\cal{L}}_\mu = {\cal{L}}_\tau$, this is obtained by
solving the integral equation
\begin{equation}
2{\cal{L}}^{\rm init} \approx {{1}\over{F_2\left(0\right)}} \int_0^{\epsilon_1}{{x^2 dx}\over{e^{x-\eta}+1}}.
\label{case1}
\end{equation}
Of course, the continuous resonance sweep scenario is an unphysical
idealization and the actual energy spectra will likely be far more complicated,
as argued above.  However, no matter the resonance sweep physics, in this
instantaneous mixing limit the numbers (number densities) of active neutrinos
in each flavor are equal and their energy spectra are identical.  In the
continuous resonance sweep scenario, in the scaled energy interval $0$ to
$\epsilon_1$, the {\it deficit} of neutrinos, $\Delta n_\nu$, is $1-2/3=1/3$ of
the original population and this deficit is identical for each active flavor.
The entire original population of {\it one} of the active neutrino flavors in
this scaled energy interval is converted to sterile neutrinos so the number
density of steriles will be $n_s=3\Delta n_\nu$ and this is related to the
initial potential lepton number through
\begin{equation}
{{n_s}\over{n_\gamma}}= {{3\Delta n_{\nu}}\over{n_\gamma}}= {{3}\over{4}}
{\cal{L}}^{\rm init}.
\label{ns1}
\end{equation}

If there is one light sterile neutrino, there may be others.  In fact it has
been claimed that two sterile species are a better fit to the LSND data than
just one \cite{conrad}. So this suggests Cases 2 and 3.

Case 2: Allow two channels of sterile neutrino production and two
kinds of light sterile neutrinos $\nu_{s1}$ and $\nu_{s2}$. As an
example, we could have $\nu_e \rightarrow \nu_{s1}$ and
$\nu_\mu\rightarrow \nu_{s2}$, but again with instantaneous mixing
among all the active neutrino flavors. A continuous resonance sweep
scenario will leave each active neutrino distribution with a low
energy hole now with $1/3$ of the normal population out to some value
of scaled neutrino energy $\epsilon_2$. In terms of the initial
potential lepton number ${\cal{L}}^{\rm
  init}={\cal{L}}_e={\cal{L}}_\mu = {\cal{L}}_\tau$, this is obtained
by solving the integral equation
\begin{equation}
{\cal{L}}^{\rm init} \approx {{1}\over{F_2\left(0\right)}}
\int_0^{\epsilon_2}{{x^2 dx}\over{e^{x-\eta}+1}}.
\label{case2}
\end{equation} 
Again, the numbers of active neutrinos in each flavor in scaled energy interval
$0$ to $\epsilon_2$ are equal and so are the deficits, $\Delta n_\nu$, which
are now $2/3$ of the original population in this interval. Now, however, the
original populations of {\it two} active flavors in this interval are converted
to sterile species so the total number density of sterile neutrinos of all
kinds is $n_s=2\cdot (3/2) \Delta n_\nu=3\Delta n_\nu$ and we have the same
relation between total sterile neutrino number density $n_s$, deficit per
flavor $\Delta n_\nu$, and initial potential lepton number as in
Eq.\ (\ref{ns1}).

For a given initial potential lepton number, this is the same total number of
sterile neutrinos (of all kinds) produced as in Case 1. However, since there
are now two channels for $\nu_s$ production, $\epsilon_2$ is smaller than
$\epsilon_1$. In Case 1, $\epsilon_1$ is relatively larger because as $\nu_e$'s
are converted to steriles two active neutrino distributions compensate by
feeding neutrinos into the hole, forcing the resonance to sweep further (higher
in energy) through the $\nu_e$ distribution to erase the net lepton numbers. In
Case 2 only one active neutrino distribution remains to compensate for the
hole.

Case 3: Allow all three active neutrinos to convert simultaneously to three
kinds of light sterile neutrinos $\nu_{s1}$, $\nu_{s2}$, $\nu_{s3}$.  Now a
smooth resonance sweep scenario will leave each active neutrino distribution
with a low energy hole with {\it zero} population out to some value of scaled
neutrino energy $\epsilon_3$. In terms of the initial potential lepton number
${\cal{L}}^{\rm init}={\cal{L}}_e={\cal{L}}_\mu = {\cal{L}}_\tau$, this is
obtained by solving the integral equation
\begin{equation}
{{2}\over{3}}{\cal{L}}^{\rm init} \approx 
{{1}\over{F_2\left(0\right)}} \int_0^{\epsilon_3}{{x^2 dx}\over{e^{x-\eta}+1}}.
\label{case3}
\end{equation} 
Now the deficits $\Delta n_\nu$ in each active flavor are equal to the original
active neutrino populations in the scaled energy interval $0$ to
$\epsilon_3$. Since all three active species transform to sterile neutrinos,
the total number density of steriles of all kinds is $n_s = 3\Delta n_\nu$. The
relation between $n_s$, $\Delta n_\nu$ and the initial potential lepton number
is the same as in Eq.\ (\ref{ns1}).  For a given ${\cal{L}}^{\rm init}$ this is
the same total number of sterile neutrinos produced as in Cases 1 and 2.
However, our three cases will have $\epsilon_1 > \epsilon_2 > \epsilon_3$ for a
given initial potential lepton number, for the reasons indicated in the last
paragraph.

For Cases 1, 2, and 3 the active neutrino distribution functions will be left
with population deficits relative to the thermal case. This is shown in
Fig.\ (\ref{figure4}) for the particular scenario where each active flavor
starts out with lepton number $L_{\nu_e}=L_{\nu_\tau}=L_{\nu_\mu}=
0.1$. Solving the above equations for the three cases yields $\epsilon_1
\approx 3.8$, $\epsilon_2 \approx 2.3$, and $\epsilon_3\approx 1.85$ in this
example.

In obvious fashion all of the above discussion applies to $\bar\nu_s$
production if the initial lepton numbers are negative.  We should also note
that the actual active and sterile neutrino energy distributions in all of the
limits considered here may differ considerably from those shown in the
figures. This is partly because the continuous resonance sweep cannot continue
to completion as described in the last section but may skip to higher scaled
energy discontinuously.  However, another source of difference from the simple
spectra shown in the figures may be because multiple neutrino mixing can be
complicated.

We have here presented a picture where the sterile neutrino undergoes a
resonance with one of the active neutrino flavors, $\nu_\alpha$.  However, if
the initial lepton numbers are equal, the resonance will instead occur between
$\nu_s$ and a superposition of the three active neutrinos.  In principle, all
three of the active neutrinos may mix with the sterile, so the MSW resonance
which is responsible for lepton number destruction may occur for the sterile
neutrino and a linear superposition of the three active neutrinos.  For
example, in the so called ``3+1'' LSND inspired mixing scheme, both
$\theta_{14}$ and $\theta_{24}$ are required to be non-zero, and the sterile
effectively mixes with all three active neutrinos (as there will also be
indirect $\nu_\tau - \nu_s$ mixing).

In any case, post-decoupling neutrino mixing cannot completely undo spectral
distortions. We conclude that a sterile neutrino in the mass range of interest
is almost certain to leave non-thermal active neutrino distribution functions
if the lepton number is significant.

\section{Constraints on Sterile Neutrinos and Lepton Numbers}

The entire plausible range of sterile neutrino masses and net lepton numbers of
interest is not likely to be consistent with all of the current observational
bounds. For example, we may demand that the initial net lepton numbers are
large enough to suppress the production of fully thermalized seas of $\nu_s$
and $\bar\nu_s$. Eq.\ (\ref{suppress}) shows that the lepton number necessary
for suppression of thermal sterile neutrino production depends both on neutrino
mass and neutrino average energy. We will hold off on considering BBN
effects/limits until the next section.

As discussed in the introduction, a population of sterile neutrinos could
provide a sufficiently large contribution to the dark matter density, depending
again on sterile neutrino mass, to run afoul of large scale structure/CMB
bounds \cite{murayama}. The above-cited analysis of the SDSS data \cite{Sloan}
using CMB anisotropy limits, galaxy clustering and bias, and coupled with the
matter power spectrum inferred from the Lyman-alpha forest suggest a limit on
the neutrino mass of $0.79\,{\rm eV}\ ({\rm 95\%\, CL})$. This corresponds to a
limit on the neutrino closure fraction
\begin{equation}
\Omega^{\rm lim}_\nu h^2 < 0.0084 \ ({\rm 95\%\, CL}),
\label{numasslim}
\end{equation}
where $h$ is the Hubble parameter at the current epoch in units of
$100\,{\rm km}\,{\rm s}^{-1}\,{\rm Mpc}^{-1}$.  This is comparable to
the older Wilkinson Microwave Anisotropy Probe (WMAP) bound,
$\Omega^{\rm lim}_\nu h^2 < 0.0076\ ({\rm 95\%\, CL})$. However, the
Eq.~(\ref{numasslim}) bound is more appropriate here as it assumes a
\lq\lq 3+1\rq\rq\ neutrino mass scenario in contrast to the three
neutrinos with a common mass assumed in the WMAP analysis. Of course,
this mass limit and our inferred limit on the closure fraction are
approximate because our sterile neutrinos have nonthermal energy
spectra.  Adopting the Eq.~(\ref{numasslim}) bound suggests that
thermal distributions of $\nu_\alpha$ and $\bar\nu_\alpha$ neutrinos
are acceptable only if they have rest masses
 \begin{equation}
m_{\nu_\alpha} \lesssim 0.79\,{\rm eV} 
{\left[ {{2F_2\left( 0\right)}\over{F_2\left( \eta_{\nu_\alpha}\right)+F_2\left(-\eta_{\nu_\alpha} \right) }} \right]} 
{\left[ {{\Omega_\nu^{\rm lim} h^2}\over{0.0084}} \right]} , 
 \label{nuslim}
 \end{equation} 
where $\alpha ={\rm e},\mu,\tau,{\rm s}$.  We can connect this with the a
putative thermal sterile neutrino sea by noting that $m_{\nu_s} \approx \left(
\delta m^2_{\rm as} \right)^{1/2}$. So, for example, $\delta m^2_{\rm as} >
0.63\,{\rm eV}^2$ is disallowed if all the sterile neutrino species have
thermal distributions. This would eliminate much of the LSND-inspired sterile
neutrino mass range.
 
However, the coherent sterile neutrino production scenarios discussed above may
do better at creeping in under the closure contribution bound. For one thing,
only $\nu_s$ (or $\bar\nu_s$) and not its opposite helicity partner are
produced coherently. Furthermore, the sterile neutrinos are produced in numbers
of order the initial lepton number. This will be smaller than a general thermal
population.

\begin{figure}
\includegraphics[width=3.25in]{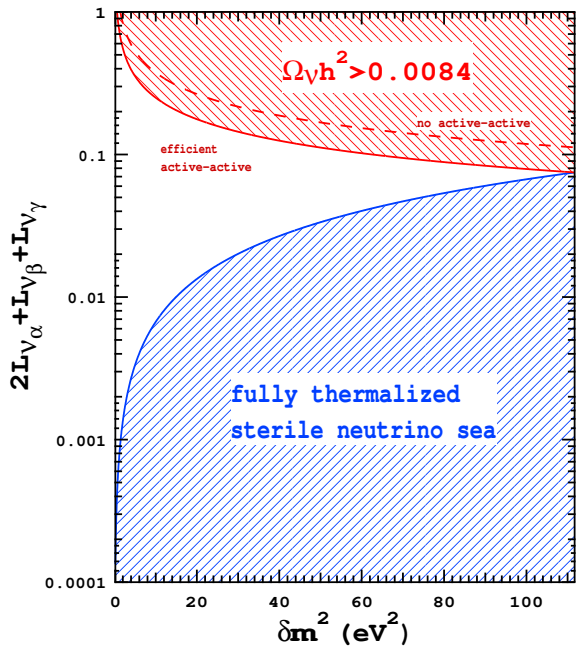}
\caption{Constraints on the the ranges of active-sterile mass-squared splitting
  and potential lepton number as derived in the smooth and continuous resonance
  sweep limit. Parameter ranges which give sufficient sterile neutrino
  production to exceed the bound on the neutrino closure fraction are shown
  cross hatched, as are parameter ranges which allow for complete or nearly
  complete thermal, undiluted energy distribution functions for a sterile
  species.  The upper solid line is for efficient active-active mixing in Cases
  1, 2 , or 3, while the upper dashed line gives the constraint for $\nu_\alpha
  \rightarrow \nu_s$ with no active-active mixing. }
\label{figure5}
\end{figure}

At the epoch of coherent sterile neutrino production the ratio of the number of
active neutrinos $\Delta n_{\nu_\alpha}$ converted to steriles to the number
total density of a thermal distribution of $\nu_\alpha$ plus $\bar\nu_\alpha$
neutrinos is in the ratio of the closure contributions of a sterile species to
thermal neutrino species:
\begin{equation}
{{\Omega_s h^2}\over{\Omega^{\rm therm}_{\nu_\alpha +\bar\nu_\alpha} h^2  }}   \approx  R_s\equiv
 {{N_s \Delta n_{\nu_\alpha}}\over{ n_{\nu_\alpha}+n_{\bar\nu_\alpha} }},
\label{xlim1}
\end{equation}
where $N_s$ is the number of active-sterile mixing channels operating in the
production of sterile neutrinos.  In turn it can be shown that in the
continuous resonance sweep scenario
\begin{equation}
R_s
 \approx  
 {\left[ {{1}\over{F_2\left( 0\right)}} \int_0^{\epsilon_{\rm c.o.}}
 {{{x^2\ dx}\over{e^{x-\eta}+1}} } \right]} 
{\left[ {{F_2\left( 0\right)}\over{F_2\left( \eta_{\nu_\alpha}\right)+F_2\left(-\eta_{\nu_\alpha} \right) }} \right]}
\label{xlim2}
\end{equation}
where $\epsilon_{\rm c.o.}$ and the degeneracy parameter $\eta$ are values
consistent with the particular sterile neutrino production scheme. From these
relations we can show that
\begin{equation}
\Omega_s h^2 \approx \left(1.062\times{10}^{-2} \right) \left( {{\beta}\over{2}}\right)
{\cal{L}} {\left[ {{\delta m^2_{\rm as}}\over{{\rm eV}^2}} \right]}^{1/2},
\label{omeganus}
\end{equation}
where ${\cal{L}}$ is an appropriate potential lepton number and where $\beta$
is a parameter that is related to the particular sterile neutrino production
scheme and the number of active-sterile channels in that scheme. For example,
$\beta = 2$ for Cases 1, 2, and 3 of the efficient active-active limit, whereas
$\beta = 4/3$ for $\nu_\alpha \rightarrow\nu_s$ only with no active-active
mixing. All of these constraints are summarized in Fig.~(\ref{figure5}).
 
Though we have employed the continuous resonance sweep scenario, the
constraints shown stem from closure fraction considerations and,
therefore, depend principally on numbers of sterile neutrinos produced
and on their rest mass. The number of sterile neutrinos is tied to the
initial potential lepton number and is insensitive to the details of
resonance sweep physics when the all the lepton number is destroyed
and the number of neutrinos of all kinds is fixed. We expect the
latter condition to be a good approximation post Weak Decoupling.  The
former assumption is a good one over most of the range of neutrino
masses/mixing considered here because neutrino flavor evolution is
highly adiabatic for these parameters as shown above.

However, the actual energy spectra of the sterile neutrinos may come into play
as precision in observations of the matter fluctuation power spectrum
increase. Current constraints are most strongly dependent on the suppression of
small scale power.  Future constraints may be able to constrain the
collisionless damping scale of neutrinos---at large scales---much more
strongly, and are most important in regard to spectral constraints.  The
collisionless damping scale for neutrinos is essentially their free streaming
length, which of course depends on the neutrino velocities which, in turn,
depend on the neutrino energy spectra.

\section{Non-Thermal Neutrino Energy Spectra: Alteration of the Relationship 
between Lepton Numbers and Primordial $^4${He} Abundance}

Primordial nucleosynthesis is a freeze out from Nuclear Statistical Equilibrium
(NSE). In NSE the abundance of nuclei is set by a competition between disorder
(entropy) and binding energy.  In the early universe alpha particles win this
competition because the entropy per baryon is very high and alpha particles
have a binding per nucleon not terribly different from iron.  At a temperature
$T_{\alpha}\sim 100\,{\rm keV}$, alpha particles form aggressively,
incorporating essentially all neutrons (all but $\sim 1/10^5$).  Therefore, the
primordial $^4${He} yield is determined roughly by the neutron-to-proton ratio
$n/p$ at $T_{\alpha}$. In mass fraction, this is $X_\alpha \approx
2{\left(n/p\right)}/{\left(n/p+1\right)}$, or $25\%$ for $n/p=1/7$. The
standard BBN $^4$He mass fraction yield prediction is $(24.85\pm 0.05)\%$ using
the CMB (Cosmic Microwave Background) anisotropy-determined baryon density
\cite{cyburt}. (The baryon closure fraction as derived from the deuterium
abundance \cite{OMeara} is consistent with the CMB-derived value.)

The observationally inferred primordial helium abundance has a long and
troubled history.  One group pegs this abundance at $0.238\pm 0.002\pm 0.005$
\cite{Olive}, while another using similar but not identical compact blue galaxy
data estimates $0.2421\pm 0.0021$ \cite{IT}.  These values are quite
restrictive. However, these older estimates may now be superseded by more
recent analyses as discussed in the Introduction.  In particular, a more
detailed analysis of the helium and hydrogen emission lines done in
Ref. \cite{OS} suggests that the allowable range of mass fraction for
primordial $^4$He is $0.232$ to $0.258$.  This is fairly generous compared to
previous \lq\lq limits,\rq\rq\ but it is still a good bet that a $5\%$ or
$10\%$ increase in the calculated, {\it predicted} yield in $^4${He} would be
an unwelcome development.

The relationship between neutrino physics and/or lepton numbers and the
primordial helium abundance remains a cornerstone of modern cosmology.
Distortions in the $\nu_e$ and $\bar\nu_e$ energy spectra stemming from
active-sterile conversion can alter lepton capture rates on nucleons and
thereby change $n/p$ and the $^4$He yield compared to a standard BBN scenario
with thermal neutrino energy spectra.

The neutron-to-proton ratio is set by the competition of the expansion rate of
the universe and the rates of the following lepton capture/decay processes:
\begin{equation}
\nu_e+n\rightleftharpoons p+e^-,
\label{nuen}
\end{equation}
\begin{equation}
\bar\nu_e+p\rightleftharpoons n+e^+,
\label{nuebarp}
\end{equation}
\begin{equation}
n\rightleftharpoons p+e^-+\bar\nu_e.
\label{ndecay}
\end{equation}
We denote the forward and reverse rates of the first process as $\lambda_{\nu_e
  n}$ and $\lambda_{e^- p}$, respectively. Likewise, the forward and reverse
rates of the second process are $\lambda_{\bar\nu_e p}$ and $\lambda_{e^+ n}$,
respectively, while those of the third process are $\lambda_{n\ {\rm decay}}$
and $\lambda_{pe\bar\nu_e}$, respectively.

The lepton capture processes' influence on the isospin state of nucleons can be
appreciable, even for the post Weak Decoupling epoch.  This is because the
number densities of relativistic neutrinos and charged leptons are some $10$
orders of magnitude larger than the baryon density.  At high enough temperature
($T \gg 1\,{\rm MeV}$), where these rates are very fast, the isospin of any
nucleon will flip from neutron to proton and back at a rate which is rapid
compared to the expansion rate of the universe, establishing a steady state
equilibrium.

As the universe expands and the temperature drops the rates of the lepton
capture processes will drop off quickly. Eventually the lepton capture rates
will fall below the expansion rate and $n/p$ will be frozen in, save for free
neutron decay. Traditionally, this \lq\lq weak freeze-out\rq\rq\ epoch is taken
to be $T_{\rm wfo} \approx 0.7\,{\rm MeV}$.

However, there is no sharp freeze-out of isospin. In fact, the
neutron-to-proton ratio $n/p$ is modified by the lepton capture reactions down
to temperatures of several hundred keV and by neutron decay through the epoch
of alpha particle formation $T_\alpha$.

The evolution of the electron fraction $Y_e=1/(1+n/p)$ throughout the expansion
is governed by
\begin{equation}
{{dY_e}\over{dt}}=\Lambda_n-Y_e\, \Lambda_{\rm tot},
\label{dyedt}
\end{equation} 
where the sum of the rates of the neutron destroying processes is
$\Lambda_n \equiv \lambda_{\nu_e n} +\lambda_{e^+ n} +\lambda_{n{\rm decay}}$
and the sum of all weak isospin changing rates is $\Lambda_{\rm tot} \equiv
\Lambda_n + \lambda_{\bar\nu_e p} +\lambda_{e^- p}+\lambda_{pe\bar\nu_e}$.

In the limit where the isospin flip rate is
fast compared to the expansion rate $H$,
the neutron-to-proton ratio has a steady state equilibrium
value
given by \cite{Qian}
\begin{eqnarray}
\label{ntop}
{{n}\over{p}} & = &
{{\lambda_{\bar\nu_ep}+\lambda_{e^-p}+\lambda_{pe\bar\nu_e}}\over
{\lambda_{\nu_en}+\lambda_{e^+n}+\lambda_{n\ {\rm
decay}}}}, \\ & \approx &
{{\lambda_{\bar\nu_ep}+\lambda_{e^-p}}\over {\lambda_{\nu_en}+\lambda_{e^+n}}}
. \nonumber
\end{eqnarray}
This solution corresponds to the fixed point ${{dY_e}/{dt}}=0$ in
Eq.~(\ref{dyedt}), where $Y_e = \Lambda_n/\Lambda_{\rm tot}$.  The second
approximation in Eq.\ (\ref{ntop}) is valid at temperatures high enough that
the rates of the three body processes can be neglected relative to the lepton
capture rates.

\begin{figure}
\includegraphics[width=3.25in]{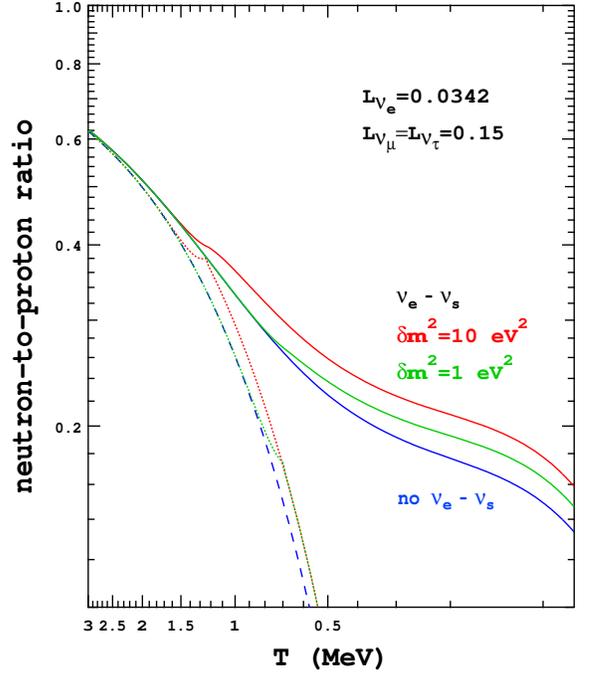}
\caption{The neutron to proton ratio as a function of temperature for the
  indicated lepton numbers. The lower solid curve gives the case with lepton
  numbers alone with no active-sterile neutrino conversion.  The dashed line is
  the equilibrium prediction for the $n/p$ ratio in this case.  The upper solid
  curve is the $n/p$ ratio for the same lepton numbers, but now with
  active-sterile conversion in the $\nu_e \rightarrow \nu_s$ channel with
  $\delta m^2=10\,{\rm eV}^2$, while the middle solid curve is for $\delta m^2
  =1\,{\rm eV}^2$.  The upper dotted curve is the steady state equilibrium
  prediction for the $n/p$ ratio for the $\delta m^2=10\,{\rm eV}^2$ case,
  while the lower dotted curve gives the same prediction for the $\delta
  m^2=1\,{\rm eV}^2$ case.}
\label{figure6a}
\end{figure}

As the universe expands and the temperature drops, the relative values of these
rates change and, hence, so does the neutron-to-proton ratio. Certainly at
temperatures $T> T_{\rm wfo}$ the three body lepton capture and free neutron
decay processes have rates which are unimportant compared to those of the
lepton capture rates. For temperatures $T\gg T_{\rm wfo}$, typical lepton
energies are large compared to the energy thresholds in the forward rate of the
process in Eq.\ (\ref{nuebarp}) and the reverse rate for the process in
Eq.\ (\ref{nuen}), so that if the lepton numbers are small we would have $n/p
\approx 1$, or $Y_e \approx 1/2$.  During the epoch $T_{\rm wfo}>T>T_\alpha$
the lepton capture processes gradually give way to free neutron decay as the
principal $n/p$-altering mechanism.  Over this time period $e^\pm$-annihilation
heats the photon/electron plasma relative to the neutrinos, further altering
the weak rates (including the free neutron decay) by modifying the neutrino and
antineutrino distribution functions relative to those for $e^\pm$. (See
Ref. \cite{Smith:1992yy} for a discussion of primordial nucleosynthesis.)

If the electron neutrinos and antineutrinos and the electrons and positrons all
have Fermi-Dirac energy spectra, then Eq.\ (\ref{ntop}) can be reduced to
\cite{CF96,FFN4}
\begin{equation}
{{n}\over{p}} \approx {{\left(\lambda_{e^-p}/\lambda_{e^+n}\right)+e^{-\eta_{\nu_e}+\eta_e-\xi}}\over
{\left(\lambda_{e^-p}/\lambda_{e^+n}\right) e^{\eta_{\nu_e}-\eta_e+\xi}+1}},
\label{thernalnp}
\end{equation}
where $\eta_e=\mu_e/T$ is the electron degeneracy parameter and $\xi =\left(
m_n-m_p\right)/T\equiv \delta m_{np}/T\approx {1.293\,{\rm MeV}/T}$ is the
neutron-proton mass difference divided by temperature.  Eq.~(\ref{thernalnp})
is approximate because it assumes identical neutrino and plasma
temperatures. We have also neglected the neutron decay/three-body capture
processes of Eq.\ (\ref{ndecay}). The expression in Eq.\ (\ref{thernalnp}) is
generally true for Fermi-Dirac leptonic energy distribution functions, even if
the neutrinos and electrons/positrons are not in true thermal and chemical
equilibrium. If and only if chemical equilibrium {\it actually} obtains (or did
obtain at some early epoch) are we guaranteed to have
$\mu_e-\mu_{\nu_e}=\mu_n-\mu_p$, where $\mu_n$ and $\mu_p$ are the neutron and
proton total chemical potentials, respectively, and only in this case does
Eq.\ (\ref{thernalnp}) reduce to
\begin{equation}
{{n}\over{p}} \approx e^{\left({\mu_e-\mu_{\nu_e}-\delta m_{np}}\right)/{T}}.
\label{chemnp}
\end{equation}
With strict chemical equilibrium and with Fermi-Dirac energy distributions for
all leptons, we could conclude from Eq.\ (\ref{chemnp}), for example, that a
positive chemical potential for electron neutrinos ({\it i.e.}, an excess of
$\nu_e$ over $\bar\nu_e$) would suppress the steady state equilibrium
neutron-to-proton ratio relative to that for $\eta_{\nu_e} =0$. This behaviour
follows also from a straightforward application of Le Chatlier's principle to
the processes in Eqs.~(\ref{nuen}) and (\ref{nuebarp}). A decrease in the
neutron abundance translates, in turn, into a decrease in the predicted
$^4${He} yield.

However, if the neutrino distribution functions are modified by active-sterile
neutrino conversion, $\nu_\alpha\rightleftharpoons\nu_s$, then the resulting
active and sterile neutrino distribution functions would not be Fermi-Dirac in
character and we could not employ Eq.\ (\ref{chemnp}) to determine the
neutron-to-protion ratio in steady state equilibrium.  Instead, we would be
forced in this case to evaluate and follow the rates directly.

We solve Eq.~(\ref{dyedt}) numerically, assuming a homogeneous and isotropic
FLRW universe. In these calculations we take the co-moving entropy density to
be conserved and thereby calculate self consistently the $e^\pm$ densities and
the neutrino and plasma (photon/$e^\pm$) temperatures. All neutrino energy
densities are handled correctly for all assumed lepton numbers and sterile
neutrino populations and, therefore, the expansion rate is also self
consistently calculated.  At each time step in these computations we employ
appropriate neutrino, antineutrino and $e^\pm$ distribution functions and
calculate the lepton capture rates $\lambda_{\nu_e n}$, $\lambda_{e^- p}$,
$\lambda_{\bar\nu_e p}$, $\lambda_{e^+ n}$, and the free neutron decay rate
$\lambda_{n {\rm decay}}$. For convenience we adjust the Coulomb wave
correction factor (see Appendix A) for the free neutron decay to be about
$\langle G\rangle \approx 1.0227$ to give a vacuum (unblocked) neutron lifetime
of $888\,{\rm s}$. We employ a Coulomb wave correction $\langle G\rangle =1$
for all of our lepton capture rates and we adopt an effective $ft$-value for
all weak reactions $ft\approx {10}^{3.035}$, so that there is roughly a $\sim
2\%$ inconsistency in overall coupling between the lepton capture processes on
the one hand and free neutron decay on the other in our calculations. This is
another reason why our calculations of $n/p$ can be used only to compare trends
in various cases with and without spectral distortion and not as quantitative
nucleosynthesis predictions.

To gauge the effect of active-sterile neutrino flavor conversion on
the evolution of the $n/p$ ratio we employ a modified forced
continuous resonance sweep scenario in either the $\nu_e\rightarrow
\nu_s$ or $\bar\nu_e\rightarrow \bar\nu_s$ channels. We use
Eq.~(\ref{ereseqn}) but fix the potential lepton number at half the
initial value. This will leave a low energy \lq\lq hole\rq\rq\ with
zero $\nu_e$ (or $\bar\nu_e$) population, as in Fig.~(\ref{figure1}),
that will grow with the expansion of the universe until $\epsilon_{\rm
  c.o.}$ is reached. The final high energy edge of the \lq\lq
hole,\rq\rq\ $\epsilon_{\rm c.o.}$ is determined from the initial
potential lepton number as discussed in the previous sections.

The modification of the lepton capture rates in the forced continuous resonance
sweep scenario is discussed in Appendix A. Consider, {\it e.g.}, $\nu_e
\rightarrow\nu_s$. Because in this case there are now {\it fewer} $\nu_e$'s,
electron capture on protons will be less Fermi blocked and, hence, the capture
rate, $\lambda_{e^-p}$, will be larger. By the same token, fewer $\nu_e$'s will
translate into a reduction of the $\nu_e$ capture rate on neutrons,
$\lambda_{\nu_en}$.  Note that both a larger value for $\lambda_{e^-p}$ and a
smaller value for $\lambda_{\nu_en}$ go in the direction of {\it increasing}
the neutron-to-proton ratio in weak steady state equilibrium. This is obvious
from Eq.\ (\ref{ntop}). Likewise, this trend in the rates will similarly make
itself felt in the general solution of Eq.~(\ref{dyedt}) and the net result
will be an increase in $n/p$ at $T_\alpha$.
 
Of course, we argued above that the actual neutrino or antineutrino spectral
distortions are likely to be quite different from the simple ones that
continuous resonance sweep would give. Since we cannot calculate the actual
neutrino spectra, we cannot give a quantitative calculation of nucleosynthesis
yields. We seek here to give rough guidelines as to where one might expect
significant neutrino spectral distortion modification of helium yields.

The forced continuous resonance sweep picture will serve to get the general
features of the rate effects. Furthermore, it {\it sometimes} will do this in a
conservative manner, especially for low lepton numbers where $\epsilon_{\rm
  c.o.}$ is small, {\it i.e.}, $\epsilon_{\rm c.o.} < 3$. As shown in Appendix
A, the weak cross sections weight the square of the neutrino energy.  A more
realistic \lq\lq picket fence\rq\rq\ neutrino spectrum will remove population
at higher neutrino energies than will the continuous sweep cut-off spectrum and
will, therefore, sometimes result in larger suppression of neutrino capture
rates and increases in $e^\pm$ capture rates. Likewise, active-active mixing
will partially fill in the \lq\lq hole\rq\rq\ in the neutrino spectra, but at
the cost of pushing the neutrino spectral deficit (relative to a thermal
spectrum) to higher energy.

Let us consider a particular active-sterile neutrino conversion
scenario in the channel $\nu_e \rightarrow \nu_s$. In this example we
take each of the three neutrino flavors to have a lepton number near
or at the maximum allowed without spectral distortion. We take
$L_{\nu_\mu}=L_{\nu_\tau}=0.15$, corresponding to degeneracy
parameters $\eta_{\nu_\mu}=\eta_{\nu_\tau}\approx 0.219$, and take
$L_{\nu_e} \approx 0.0343$, corresponding to electron neutrino
degeneracy parameter $\eta_{\nu_e}=0.05$. This will give an initial
potential lepton number in the $\nu_e\rightarrow\nu_s$ transformation
channel,
\begin{equation}
\label{potleptest}
{\cal{L}}_e^{\rm initial} = 2L_{\nu_e}+L_{\nu_\mu}+L_{\nu_\tau}\approx
0.368.
\end{equation} 
In this case the difference in neutrino energy density over the zero lepton
case is only $\sim 0.2\%$. Therefore, the expansion rate of the universe at Weak
Freeze Out in this case will differ from the standard BBN model by only $\sim
0.2\%$. Therefore, the expansion rate {\it by itself} would give a negligible
difference in neutron-to-proton ratio between the case with the lepton number
in Eq.~(\ref{potleptest}) and the zero lepton number, standard BBN case.

Nevertheless, we calculate the $n/p$ ratio completely self
consistently as described above, beginning with steady state
equilibrium at $T=3\,{\rm MeV}$. The results are shown in
Fig.~(\ref{figure6a}).  The lower solid curve is the $n/p$ ratio in
the case with the lepton numbers but with no active-sterile neutrino
flavor transformation and, therefore, no sterile neutrino population
and no active neutrino spectral distortion. As expected, the $n/p$
ratio is suppressed relative to a zero-lepton number standard BBN
calculation owing to the large positive $\nu_e$ degeneracy parameter
$\eta_{\nu_e}=0.05$. At the lowest temperature on the plot, $T\approx
80\,{\rm keV}$ (this is near $T_{\alpha}$ in an actual nucleosynthesis
calculation for baryon number $\eta = 6\times {10}^{-10}$) we have
$n/p\approx 0.133$, corresponding to a rough $^4$He yield $X_\alpha
\approx 23.5\%$. A similar calculation but with zero lepton numbers
(this is the standard cosmological model case) gives $n/p\approx
0.141$, corresponding to $X_\alpha^0\approx 24.7\%$, so our
lepton number-only case corresponds to about a $5\%$ decrease in
helium yield, as expected.

The dashed line on Fig.~(\ref{figure6a}) is the equilibrium $n/p$ ratio given
by Eq.~(\ref{thernalnp}) or Eq.~(\ref{chemnp}) (they give the same result in
this case because all leptons have Fermi-Dirac distribution functions). We see
that the lepton capture rates become too slow to maintain equilibrium when $T <
2\,{\rm MeV}$, though the actual $n/p$ ratio is still significantly influenced
by these rates for temperatures greater than $T \sim 300\,{\rm keV}$.

The upper solid curve in Fig.~(\ref{figure6a}) is the $n/p$ ratio in the case
where the same lepton numbers now drive $\nu_e\rightarrow \nu_s$ in the
modified forced continuous resonance sweep case as described above with $\delta
m^2 = 10\,{\rm eV}^2$. In this scenario the resonance will, eventually, sweep
out to $\epsilon_{\rm c.o.} \approx 2.724$. The upper dotted curve is the
steady state equilibrium $n/p$ ratio for this case, given by
Eq.~(\ref{thernalnp}). The \lq\lq kink\rq\rq\ in this curve (mirrored in a
similar, smaller deviation in the solid curve) at about $T\approx 1.2\,{\rm
  MeV}$ corresponds to the point where for this $\delta m^2$ and ($1/2$)
potential lepton number the resonance has swept far enough ({\it i.e.,} near
$\epsilon_{\rm c.o.}$, see Eq.~\ref{eresapprox}) to significantly reduce the
$\nu_e+n\rightarrow p+e^-$ rate.  From then on the $n/p$ ratio tracks higher
than the lepton number-only case and at $T\approx 80\,{\rm keV}$ we have
$n/p\approx 0.159$, corresponding to $X_\alpha \approx 27.4\%$, an $11\%$
increase over the standard BBN zero-lepton number case, and a whopping nearly
$17\%$ increase over the lepton number-only case. (In fact, the bigger neutron
number density in this case likely would lead to a slightly earlier assembly of
alpha particles, {\it i.e.}, a higher $T_\alpha$ and, hence, an even slightly
bigger $^4$He yield.)

In other words, the existence of a sterile neutrino that mixes with the $\nu_e$
completely altered the {\it sign} of the effect of a net lepton number. The
lepton number by itself would have given a comfortable {\it reduction} of the
helium yield, whereas the $\nu_e$ spectral distortion in this case
re-engineered this into an uncomfortable {\it increase} in the helium over the
standard model. Despite the limitations of our calculations and approximations
it is clear from this example that the existence of sterile neutrinos could
alter the relationship between lepton number(s) and $^4$He yield.

Note that there would be significant alterations in the
lepton number/helium-yield relation even if the resonance did not sweep beyond
$\epsilon_{\rm max}$. We have already argued that the resonance will find a way
to sweep beyond $\epsilon_{\rm max}$, given that the resonance condition can be
met for non-continuous resonance sweep and that neutrino flavor evolution is
likely quite adiabatic for the relevant conditions. However, for argument's
sake, let us adopt the same example as above and with $\delta m^2=10\,{\rm
  eV}^2$, but now limit the resonance's progress to $\epsilon_{\rm max}$. For
this example ($L_{\nu_\mu}=L_{\nu_\tau}=0.15$ and $L_{\nu_e} \approx 0.0343$)
we have $\epsilon_{\rm max} \approx 1.461$. The numerical $Y_e$ calculation in
this case, using the same scenario as above, yields $n/p\approx 0.138$ at
$T\approx 80\,{\rm keV}$, a nearly $4\%$ increase over the lepton number-only
case.

Our spectral distortion effects can be dependent on $\delta m^2$, at least for
the high lepton numbers adopted in the above example
($L_{\nu_\mu}=L_{\nu_\tau}=0.15$ and $L_{\nu_e} \approx 0.0343$). The middle
solid curve in Fig.~(\ref{figure6a}) is the $n/p$ ratio evolution for this case
with forced continuous resonance sweep out to $\epsilon_{\rm c.o.}$, but now
for $\delta m^2=1\,{\rm eV}^2$. The lower dotted curve is the
Eq.~(\ref{thernalnp}) steady state equilibrium $n/p$ for this case. Since the
resonance sweep is proportional to $\delta m^2$ (see Eq.~\ref{eresapprox}), we
expect that the resonance sweep will not have progressed far enough to decrease
the $\nu_e$ capture rate significantly until a lower temperature than in the
$\delta m^2=10\,{\rm eV}^2$ case. That temperature is about $T\approx 700\,{\rm
  keV}$ in this case. The result is that the rate reductions come in later
here, where they are less effective at lowering $n/p$, though there is still a
hefty effect. At $T\approx 80\,{\rm keV}$ we have $n/p\approx 0.147$,
corresponding to $X_\alpha \approx 25.6\%$, a nearly $4\%$ increase over our
standard BBN zero-lepton number case, and a $9\%$ increase over the
lepton number-only case. If we do the same calculations but now for $\delta
m^2=0.2\,{\rm eV}^2$ we get $X_\alpha \approx 24.7\%$, the same as the zero
lepton number case and a $5\%$ increase over the lepton number-only case, again
a significant but smaller effect.

However, we do not see this level of $\delta m^2$ dependence in $n/p$
alterations when the lepton numbers are {\it small}. This is because
$\epsilon_{\rm res} \sim \delta m^2/{\cal{L}}$, so a low ${\cal{L}}$ translates
into more progress in resonance sweep for a given $T$ and $\delta m^2$.

A case in point is where $L_{\nu_e}=L_{\nu_\mu}=L_{\nu_\tau}=0.01$ ($\nu_e$
degeneracy parameter $\eta_{\nu_e} \approx 0.0146$), corresponding to potential
lepton number ${\cal{L}}_e=0.04$ for the $\nu_e\rightarrow\nu_s$ channel, with
$\epsilon_{\rm c.o.}\approx 0.96$. This equilibrated case represents a
threshold: values of ${\cal{L}}_e$ larger than this in fully equilibrated limit
could produce significant ($> 1\%$) increases in helium yield over the
lepton number-only scenario, depending on the resonance sweep scenario and the
efficacy of active-active transformation and residual neutrino down-scattering.
The lepton number-only calculation for this case (as in Fig.~\ref{figure6a}) at
$T\approx 79\,{\rm keV}$ gives $n/p\approx 0.139$, or $X_\alpha \approx
24.4\%$, a very slight decrease from the standard zero lepton number case
discussed above. However, with $\delta m^2 \ge 0.2\,{\rm eV^2}$ and
$\nu_e\rightleftharpoons\nu_s$ conversion enabled, at $T\approx 79\,{\rm keV}$
we get $n/p\approx 0.14$, or $X_\alpha \approx 24.6$, a $\sim 1\%$ increase
over the lepton number-only case. Larger values of ${\cal{L}}_e$ will give
bigger discrepancies between the cases with and without sterile neutrino mixing
in the forced continuous sweep limit, and may well do so in more realistic
resonance sweep scenarios as well.

Likewise, we can investigate the analogous limit for un-equilibrated cases
(where active-active mixing is ineffective) by employing $n/p$ evolution
calculations along the lines of those presented in Fig.~(\ref{figure6a}). We
find that the case with $L_{\nu_e}=0.001$ and $L_{\nu_\mu}=L_{\nu_\tau}=0.01$
(corresponding to ${\cal{L}}_e=0.022$, $\epsilon_{\rm c.o.}\approx 0.76$ and
$\eta_{\nu_e}\approx 0.00146$) is likely to give about a $1\%$ increase in
$X_\alpha$ over the lepton number-only case in the forced continuous resonance
sweep scenario. Again, depending on the actual resonance sweep history and the
efficacy of neutrino scattering this represents a warning point for the
un-equilibrated cases.

With these numerical calculations we are led to ask the following question: at
what point are the {\it positive} lepton numbers big enough that when combined
with a sterile neutrino and concomitant spectral distortions we {\it exceed}
the classic observationally inferred helium limits discussed above
($X_\alpha^{\rm lim} \approx 25\%$)? In the forced continuous resonance sweep
scenario, and taking active-active mixing as efficient, we find that limit
occurs for equilibrated lepton numbers near
$L_{\nu_e}=L_{\nu_\mu}=L_{\nu_\tau}=0.09$ ($\nu_e$ degeneracy parameter
$\eta_{\nu_e} \approx 0.131$), corresponding to potential lepton number
${\cal{L}}_e=0.36$ for the $\nu_e\rightarrow\nu_s$ channel, with $\epsilon_{\rm
  c.o.}\approx 2.57$. The lepton number-only calculation for these parameters
gives $n/p\approx 0.121$ or $X_\alpha\approx 21.6\%$ at $T\approx 79\,{\rm
  keV}$; whereas, active-sterile transformation in this case with $\delta
m^2=10\,{\rm eV}^2$ yields at this temperature $n/p\approx 0.143$ or $X_\alpha
\approx 25\%$. However, this drops to $X_\alpha \approx 24.6\%$ for $\delta
m^2=5\,{\rm eV}^2$ and is down to $X_\alpha\approx 23\%$ for $\delta
m^2=0.2\,{\rm eV}^2$. For these lepton number parameters we note that $\delta
m^2> 5\,{\rm eV}^2$ may already be ruled out by closure constraints.

\begin{figure}
\includegraphics[width=3.25in]{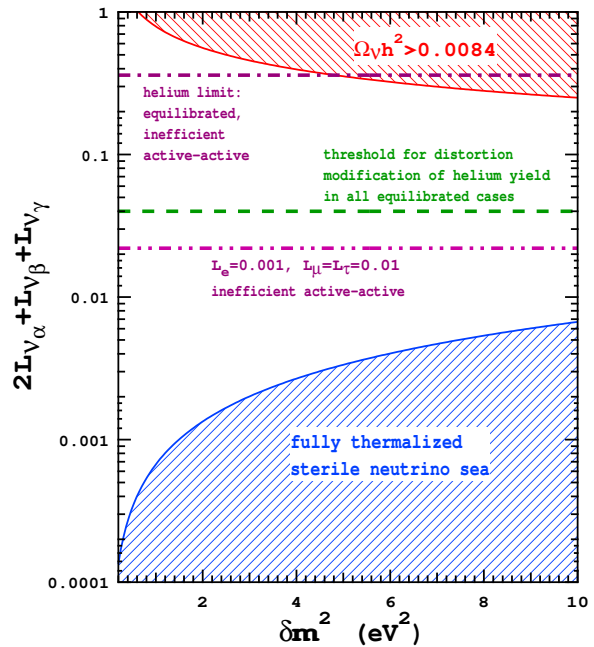}
\caption{Constraints as in Fig.~(\ref{figure5}) and now with BBN \lq\lq
  guidelines,\rq\rq\ see text. The double-dot-dashed line gives an estimate of
  the limiting $^4${He} yield in a no active-active mixing scenario where
  $L_e=0.001$ and $L_\mu=L_\tau=0.01$. The dashed line gives a threshold in
  potential lepton number for all cases where neutrino lepton numbers are
  equilibrated, including those with efficient active-active mixing. Beyond
  this threshold there could be significant alterations in the helium yield
  stemming from non-thermal neutrino distributions. The upper dash-dot line
  gives a rough estimate of the limiting potential lepton number for
  equilibrated cases with inefficient active-active mixing. }
\label{figure6}
\end{figure}

These rough cautionary guides as to where one might expect sterile
neutrino-generated spectral distortions to become important are shown
superposed on the closure fraction constraints in Fig.~(\ref{figure6}). Our
calculations are only very general guides, as discussed above, as we cannot
follow the resonance sweep physics in detail and, therefore, we cannot be
quantitative about nucleosynthesis yields.

One example of a significant deviation from the forced resonance sweep picture
is where active-active neutrino mixing is efficient and neutrino lepton numbers
equilibrate rapidly as resonance sweep occurs. In particular, in Cases 1 and 2
discussed in the last section, the neutrino populations in the \lq\lq
hole\rq\rq\ are not zero, with the consequence that for a given $\epsilon_{\rm
  c.o.}$ the rates are not as affected as in the cases discussed above. For
example, for initial lepton numbers $L_{\nu_e}=L_{\nu_\mu}=L_{\nu_\tau} =0.1$,
corresponding to $\eta_{\nu_e} \approx 0.146$, the lepton number-only case
steady state $n/p$ ratio from Eq.~({\ref{chemnp}) at $T=0.7\,{\rm MeV}$ is
  $n/p\approx 0.136$.  By contrast, when we allow sterile neutrino production
  in Case 1, the resulting non-thermal $\nu_e$ spectrum increases the steady
  state value to $n/p\approx 0.15$. In Case 2 this goes to $n/p\approx 0.145$
  while in Case 3 it is only $n/p\approx 0.144$. These examples serve to
  illustrate the change in lepton capture rates between Cases 1, 2, and 3. This
  gives a rough guide as to how spectral distortion effects on BBN might
  decrease in calculations with the full $Y_e$ evolution as we go from Case 1
  toward Case 3.

Likewise, the cases with the highest values of $\delta m^2$ coupled with lower
values of ${\cal{L}}$ in Fig.~(\ref{figure6}) may already have experienced
considerable resonant active-sterile conversion and concomitant lepton number
depletion {\it before} Weak Decoupling, while active neutrino scattering
down-scattering was still effective. This would either wash out much of the
\lq\lq hole\rq\rq\ in the neutrino distribution functions or leave a hole of
reduced energy width at higher energies.
 
What about conversion of $\bar\nu_e$ to steriles, {\it i.e.,}
$\bar\nu_e\rightarrow\bar\nu_s$? This process can be matter-enhanced when the
overall potential lepton number is negative, ${\cal{L}}_e < 0$. It will be
exactly analogous to the positive potential lepton number case, at least as far
as the neutrino flavor conversion and the resonance sweep physics goes. This
close analogy ends, however, when it comes to the lepton capture reactions.

If the resonance sweeps adiabatically in the forced resonance sweep scenario
out to a scaled antineutrino energy $\bar\epsilon_{\rm c.o.}$, a hole will be
left in the $\bar\nu_e$ distribution, in complete analogy to the cases
discussed above. Fewer $\bar\nu_e$'s will translate into a decreased
antineutrino capture rate, $\bar\nu_e+p\rightarrow n+e^+$, and an increased
rate of positron capture, $e^++n\rightarrow p+\bar\nu_e$. These rate
modifications both go in the direction of {\it decreasing} the
neutron-to-proton ratio $n/p$, as is obvious from Eq.~(\ref{ntop}).
  
One might think at first that simply changing the sign of the potential lepton
numbers given in the above examples could result in a significantly suppressed
$n/p$ at $T_\alpha$.  This is not necessarily correct however, because the
threshold in the reaction $\bar\nu_e+p\rightarrow n+e^+$ plays a crucial
role. As can be seen in the rate integrals given in Appendix A, in this channel
a $\bar\nu_e$ must have an energy in excess of the threshold,
$E_{\bar\nu_e}>E_{\bar\nu_e}^{\rm thresh}$ to be captured. The threshold is
$E_{\bar\nu_e}^{\rm thresh} = Q_{np}+m_ec^2\approx 1.804\,{\rm MeV}$.
 
In fact, in the forced continuous resonance sweep scenario, unless
$\bar\epsilon_{\rm c.o.} > E_{\bar\nu_e}^{\rm thresh}/T$, there will be {\it
  no} modifications in the $\bar\nu_e$ capture rates. Likewise for the inverse
process of positron capture on neutrons, $e^++n\rightarrow p+\bar\nu_e$. In
this reaction, there will be no alteration of the final state $\bar\nu_e$
blocking factor unless $\bar\epsilon_{\rm c.o.} > E_{\bar\nu_e}^{\rm
  thresh}/T$.

For example, consider the equilibrated case with $L_{\nu_e}
=L_{\nu_\mu}=L_{\nu_\tau}=-0.01$ ($\eta_{\nu_e}\approx -0.0146$). This gives
the opposite sign potential lepton number, ${\cal{L}}_{e}\approx -0.04$, from
the analogous positive lepton number case considered above. The forced
continuous, adiabatic resonance sweep scenario would give for this case
$\bar\epsilon_{\rm c.o.} \approx 0.96$.  A calculation of $Y_e$ with
temperature as in Fig.~(\ref{figure6a}) for the lepton number-only version of
this case gives $n/p\approx 0.148$ at $T\approx 79\,{\rm keV}$, or roughly
$X_\alpha \approx 25.7\%$, an increase in helium yield over the standard zero
lepton number case, as expected. However, when we now do the same calculation
but with $\bar\nu_e\rightarrow\bar\nu_s$ conversion with $\delta m^2>0.2\,{\rm
  eV}^2$ we obtain at the same temperature $n/p\approx 0.143$, corresponding to
$X_\alpha \approx 25\%$. This is a $\sim 3\%$ reduction over the
lepton number-only case.
 
This negative potential lepton number value again signals a threshold: negative
potential lepton numbers in the fully equilibrated limit larger in magnitude
than this could give significant modification of the relationship between
lepton number and helium yield. This guideline is shown in Fig.~(\ref{figure7})
in the same manner as for the guides for positive potential lepton numbers.
Again, the warning as to the very rough nature of our guidelines owing to
uncertain resonance sweep physics and the efficacy of neutrino down-scattering
and active-active mixing applies here as well as in Fig.~(\ref{figure6}).

\begin{figure}
\includegraphics[width=3.25in]{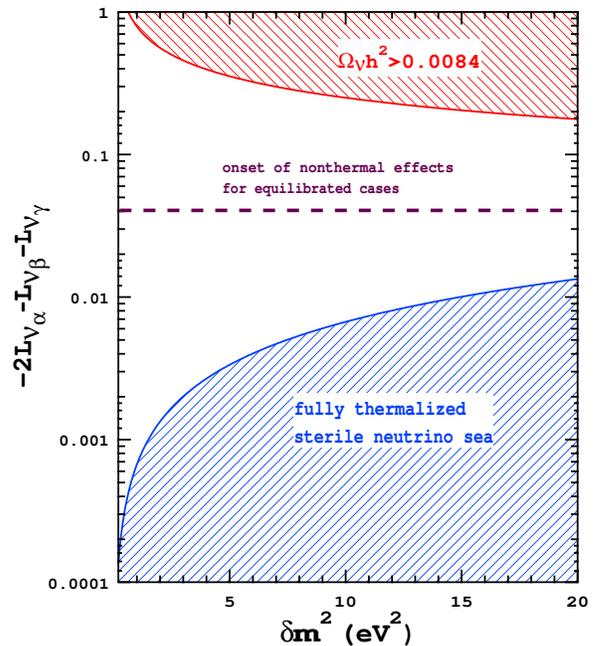}
\caption{Constraints and guidelines as in Fig.s~(\ref{figure5}) \&
  (\ref{figure6}), but now for negative values of potential lepton number.  The
  dashed line gives a rough threshold on potential lepton number in the
  equilibrated regime ( $L_e=L_\mu=L_\tau$) beyond which significant
  alterations (reductions) in $^4$He yield stemming from $\bar\nu_e$ spectral
  distortion might be expected. }
\label{figure7}
\end{figure}

Because of the threshold issue in the negative potential lepton number cases it
is not possible to find parameters that could provide a large suppression of
$^4$He over the zero lepton number case, at least in the fully equilibrated
limit. Partly this is because large lepton number magnitude would require a
large $\delta m^2$ in order to have enough resonance sweep early enough to
affect the lepton capture enough to drop $n/p$ significantly. The large $\delta
m^2$ at these large, negative ${\cal{L}}_e$ values seem to be in conflict with
closure constraints.

\section{Discussion and Conclusions}

The most general conclusions that can be drawn from this work is that
the existence of one or more light sterile neutrinos could (1) alter
the relationship between neutrino chemical potential and primordial
nucleosynthesis yields, and (2) leave both active and sterile
neutrinos with nonthermal, distorted energy spectra. Attempts to
constrain an \lq\lq LSND\rq\rq\ sterile neutrino based on conventional
degenerate primordial nucleosynthesis considerations, as well as
attempts to reconcile this neutrino with BBN limits via a primordial
lepton number are now suspect. However, obtaining the detailed
relationship between lepton numbers, active-sterile neutrino mixing
parameters and light element nucleosynthesis yields that would be
required to effect reliable constraints is beyond the scope of the
work presented in this paper.

This is mostly a consequence of another discovery made in this work: MSW
resonances cannot sweep smoothly and continuously beyond $\epsilon_{\rm
  max}$. We showed that one way the resonance condition and adiabatic
conversion criterion can be met beyond this point is if the resonance skips to
higher energies a number of times until the initial lepton number is
depleted. Partly for this reason resonance evolution is likely to be quite
complicated. Complication will also arise because active-active neutrino mixing
can be efficient and can occur simultaneously with active-sterile
transformation.

In any case, neutrino flavor conversion is almost inevitable once the resonance
condition is met. This is on account of another insight presented in this
paper: the highly adiabatic nature of neutrino flavor evolution through MSW
resonances for the neutrino mass/mixing parameters of most interest and for the
conditions in the post Weak Decoupling early universe.

If the mini-BooNE experiment sees a positive signal, confirming the existence
of light sterile neutrinos, we will be forced to confront the problems posed in
this paper. Likewise, future progress in improving the precision and confidence
in the observationally inferred primordial helium abundance coupled with CMB
and large scale structure-derived limits on neutrino collisionless damping
scales could give us hints about active-sterile neutrino mixing physics in the
early universe.

\acknowledgments    

The work of G.M.F. was supported in part by NSF grant PHY-00-99499, the TSI
collaboration's DOE SciDAC grant, and a UC/LANL CARE grant at UCSD; K.A. was
supported by Los Alamos National Laboratory (under DOE contract W-7405-ENG-36)
and also acknowledges the UC/LANL CARE grant.  N.B. was supported by Fermilab,
which is operated by URA under DOE contract No.\ DE-AC02-76CH03000, and was
additionally supported by NASA under NAG5-10842.  This research was also
supported in part by the NSF under Grant No. PHY99-07949. The authors thank the
ECT$^*$ and the INT at the University of Washington for hospitality.  We thank
Ray Volkas and Robert Foot for discussions.

\appendix
\section{Weak Rates with Non-Thermal Neutrino Energy Spectra}

In this Appendix we calculate the forward and reverse rates of the processes in
Eqs.~(\ref{nuen}) \&\ (\ref{nuebarp}) for the cases where, respectively, all
neutrinos below energy $E_{\nu_e}=T\epsilon$ or antineutrinos below energy
$E_{\bar\nu_e}=T\bar\epsilon$ are converted to sterile species. We provide
estimates of these rates in terms of standard relativistic Fermi integrals.  We
also discuss how these rates would be modified if the MSW resonance does not
sweep smoothly and continuously (and adiabatically) through the low energy
neutrino or antineutrino distribution function, but instead skips to higher
energies. The rate modifications for Cases 1 and 2 in the efficient active-active 
neutrino mixing limit will be different, of course, because in those scenarios the \lq\lq holes\rq\rq\ in the neutrino distribution functions are not empty. Though the rate formulae presented here are not valid for these cases, they still give a general idea of how the lepton capture/decay rates depend on spectral distortion and thresholds. 

If there is no active-sterile conversion and all neutrino, nucleon, and charged
lepton distribution functions are thermal in character, the $\nu_e$ capture
rate on neutrons is $\lambda_{\nu_en}^0$. By contrast, we will denote as
$\lambda_{\nu_en}$ the actual electron neutrino capture rate when the same
thermodynamic conditions obtain, but now where $\nu_e$'s have been converted to
sterile species up to scaled energy $\epsilon$ as outlined above. If all
neutrino, nucleon, and charged lepton energy distribution functions are at
least piece-wise Fermi-Dirac or zero, these rates can be written \cite{FFN4},
respectively, as
\begin{widetext}
\begin{equation}
\label{nucaprate0}
\lambda_{\nu_en}^0\approx \Lambda {\left[1-e^{\eta_e-\eta_{\nu_e}-\xi_{np}}\right]}^{-1}
\int_0^\infty{x^2{\left(x+\xi_{np}\right)}^2 {\left( {{1}\over{e^{x-\eta_{\nu_e}}+1}}-{{1}\over{e^{x+\xi_{np}-\eta_{e}}+1}} \right)} dx},
\end{equation}
\begin{eqnarray}
\label{nucaprate}
\lambda_{\nu_en} & \approx &  \Lambda {\left[1-e^{\eta_e-\eta_{\nu_e}-\xi_{np}}\right]}^{-1}
\int_{\epsilon}^\infty{x^2{\left(x+\xi_{np}\right)}^2 {\left( {{1}\over{e^{x-\eta_{\nu_e}}+1}}-{{1}\over{e^{x+\xi_{np}-\eta_{e}}+1}} \right)} dx}
\\
& \approx & \Lambda {\left[1-e^{\eta_e-\eta_{\nu_e}-\xi_{np}}\right]}^{-1} \sum_{n=0}^{4}{\alpha_n \left[F_n\left(\eta_\nu^{\rm eff}\right)-F_n\left(\eta_e^{\rm eff}\right)\right]}.\nonumber\end{eqnarray}
\end{widetext}
Here the integration variable in both equations is the scaled $\nu_e$ energy,
$x=E_{\nu_e}/T$. The final state electron energy is $E_e=T
\left(x+\xi_{np}\right)$. The other notation in these expressions is as defined
above and $\xi_{np} \equiv Q_{np}/T$ with $Q_{np}=\delta m_{np}$. There is no
threshold for $\nu_e$ energy in this reaction channel. The temperature and
matrix element-dependent factor in both rate expressions is
\begin{eqnarray}\label{Lambda}
\Lambda & \equiv & \langle G\rangle {{\ln 2}\over{\langle ft\rangle}}
{\left({{T}\over{m_e c^2}}\right)}^5 \\ & \approx & \left(1.835\times
10^{-2}\,{\rm s}^{-1}\right) \langle G\rangle {\left({{T}\over{\rm
MeV}}\right)}^5 ,\nonumber\end{eqnarray}where $\langle ft\rangle$ is the
effective $ft$-value as defined in Ref. \cite{FFN4} and is roughly $\log_{10}
ft \approx 3.035$ for free nucleons, while $\langle G\rangle$ is the average
Coulomb wave correction factor (also defined in Ref. \cite{FFN4}) with $G\equiv
F(Z,E_e) E_e/p_e$ and where $F(Z,E_e)$ is the usual Fermi function for nuclear
charge $Z$ and final state electron energy $E_e$. For the relativistic leptons
considered here (the lowest electron energy is $\approx Q_{np}\approx 1.3\,{\rm
MeV}$), $\langle G\rangle \approx 1$, though we note that $\langle G\rangle$ in
the no-transformation case is slightly larger than that for the case with the
$\epsilon$ cut-off on account of the lower energy electrons present in the
phase space integral in the former case. (Electrons are \lq\lq pulled in\rq\rq\
to the proton because of Coulomb attraction, making for a larger overlap.)

The second approximation in Eq.\ (\ref{nucaprate}) gives
$\lambda_{\nu_en}$ as a sum of differences of relativistic Fermi
integrals. In this expression the effective $\nu_e$ and $e^-$
degeneracy parameters are defined as $\eta_{\nu}^{\rm eff} \equiv
\eta_{\nu_e}-\epsilon$ and $\eta_e^{\rm eff} \equiv \eta_e-\delta$,
respectively, where $\delta \equiv \epsilon+\xi_{np}$. Also in Eq.\
(\ref{nucaprate}) we define $\alpha_4 \equiv1$, while $\alpha_3 \equiv
2\left(\epsilon+\delta\right)$, and $\alpha_2 \equiv
{\left(\epsilon+\delta\right)}^2+2\epsilon\delta$, with $\alpha_1
\equiv 2\epsilon\delta\left(\epsilon+\delta\right)$ and $\alpha_0
\equiv \epsilon^2 \delta^2$. Note that as $\epsilon\rightarrow 0$,
both expressions in Eq.\ (\ref{nucaprate}) approach $\lambda_{e^-p}^0$
in Eq.\ (\ref{nucaprate0}). It is obvious that for nonzero $\epsilon$
the $\nu_e$ capture rate on neutrons will be reduced over its
no-transformation value, $\lambda_{\nu_en} < \lambda_{\nu_en}^0$. The
rate $\lambda_{\nu_en}$ is illustrated in Fig.~\ref{nuen2} as a function of
$\epsilon$ or $L_{\nu_e}$.

\begin{figure}
\includegraphics[width=3.25in]{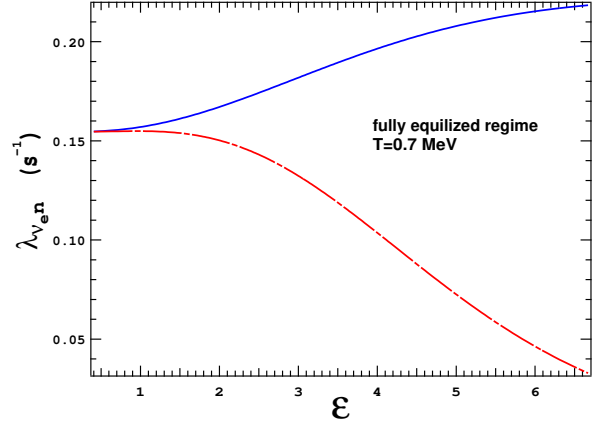}
\caption{Rate $\lambda_{\nu_e n}$ in s$^{-1}$ for the process $\nu_e+n\rightarrow p+e^-$ at temperature
$T=0.7\,{\rm MeV}$ as a function of $\epsilon$ and/or $L_{\nu_e}$ in the smooth and
continuous resonance sweep limit and for the case of complete
active neutrino equalization ($L_{\nu_e}=L_{\nu_\mu}=L_{\nu_\tau}$). The solid curve
gives the rate for no sterile neutrino conversion, thermal $\nu_e$ distribution, but with
the $\nu_e$ chemical potential appropriate for the corresponding $\epsilon$ value.
The dot-dashed curve gives the rate with active-sterile neutrino conversion and
corresponding non-thermal character for the $\nu_e$ energy
distribution function.  }
\label{nuen2}
\end{figure}

The rate for the corresponding reverse process of electron capture on protons, $e^-+p\rightarrow n+\nu_e$, will be increased if some $\nu_e$'s are transformed to sterile states, as there will be less final state $\nu_e$ blocking in this case. For a Fermi-Dirac distribution of electrons, and in terms of an integral over electron energy $E_e$, this rate is
\begin{equation}\label{genep}
\lambda_{e^-p} \approx {{\langle G\rangle \ln{2}}\over{\langle ft\rangle { \left( m_ec^2 \right)}^5  }}
\int_{Q_{np}}^\infty{{{E_e^2 \left( E_e-Q_{np} \right)^2}\over{e^{E_e/T-\eta_e}+1  }} \left[ 1-{{S}}_{\nu_e} \right] dE_e  },\end{equation}where $S_{\nu_e}$ is the energy-dependent $\nu_e$ occupation probability,
\begin{eqnarray}S_{{\nu_e}} & = & 0\ \ \ \ \ \ \ \ \ \ \ \ \ \ \ \ \ \ \ \ {\rm for}\ E_{\nu_e}/T\le \epsilon,
\\
S_{{\nu_e}} & = & {{1}\over{ e^{E_{\nu_e}/T-\eta_{\nu_e}}+1  }} \ \ {\rm for}\ E_{\nu_e}/T > \epsilon.\end{eqnarray}Here the $\nu_e$ energy is $E_{\nu_e}=E_e-Q_{np}$ on account of the threshold, $Q_{np}$.

It is convenient to re-write the rate in Eq.\ (\ref{genep}) as an integration
over neutrino energy scaled by temperature, $x=E_{\nu_e}/T$, and as a sum of
contributions from low neutrino energy with no final state blocking, and higher
final state neutrino energy where there is non-zero Fermi blocking,
\begin{equation}\label{sumrate}
\lambda_{e^-p}=\lambda_{e^-p}^{\rm low}+\lambda_{e^-p}^{\rm high}.\end{equation}The first of these rate contributions can be approximated by
\begin{equation}\label{lowrate}
\lambda_{e^-p}^{\rm low} \approx \Lambda \int_0^\epsilon{ {{x^2 \left(
x+\xi_{np} \right)^2}\over{e^{x+\xi_{np}-\eta_e}+1 }} dx}.\end{equation} Just
as for $\nu_e$ capture, the average Coulomb wave correction factor will be
lower (closer to unity) with increasing $\epsilon$. Again this has to do with
the enhancement of the low energy electron probability density near the proton.
As above, we can represent the rate contribution in Eq.\ (\ref{lowrate}) in
terms of standard relativistic Fermi integrals,
\begin{widetext}
\begin{equation}
\label{approxrate1}
\lambda_{e^-p}^{\rm low} \approx \Lambda {\left[
    F_4\left(\eta_e-\xi_{np}\right)+2\xi_{np}
    F_3\left(\eta_e-\xi_{np}\right)+\xi_{np}^2 F_2\left(\eta_e-\xi_{np}\right)
    \right]} - \Lambda \sum_{n=0}^{4}{\beta_n
  F_n\left(\eta_e-\xi_{np}-\epsilon\right)},
\end{equation}
\end{widetext}
where $\beta_4\equiv 1$, and where $\beta_3\equiv
2\left(\epsilon+\delta\right)$, while $\beta_2\equiv
{\left(\epsilon+\delta\right)}^2+2\epsilon\delta$ and
$\beta_1\equiv2\epsilon\delta{\left(\epsilon+\delta\right)}$, with
$\beta_0\equiv \epsilon^2 \delta^2$. Here we define $\delta \equiv \epsilon +
\xi_{np}$.  

\begin{figure}
\includegraphics[width=3.25in]{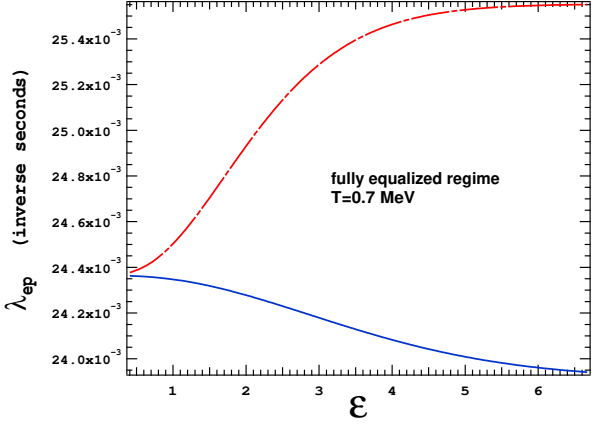}
\caption{Rate $\lambda_{ep}$ in s$^{-1}$ for the process $e^{-}+p\rightarrow n+\nu_e$
at temperature
$T=0.7\,{\rm MeV}$ as a function of $\epsilon$ and/or $L_{\nu_e}$ in the smooth and
continuous resonance sweep limit and for the case of complete
active neutrino equalization ($L_{\nu_e}=L_{\nu_\mu}=L_{\nu_\tau}$). The solid curve
gives the rate for no sterile neutrino conversion, thermal $\nu_e$ distribution, but with
the $\nu_e$ chemical potential appropriate for the corresponding $\epsilon$ value.
The dot-dashed curve gives the rate with active-sterile neutrino conversion and
corresponding non-thermal character of the $\nu_e$ energy distribution function.  }
\label{eleccap2}
\end{figure}

The physical interpretation of this expression for $\lambda_{e^-p}^{\rm low}$
is clear if it is recalled that the $\nu_e$ energy is $E_{\nu_e}=E_e-Q_{np}$,
implying that the \lq\lq effective final state neutrino degeneracy
parameter\rq\rq\ is $\eta_e-\xi_{np}$ for the no-conversion case, and
$\eta_e-\xi_{np}-\epsilon$ with conversion of $\nu_e$'s to steriles. Of course,
as $\epsilon\rightarrow 0$, the rate contribution from the (final state
$\nu_e$) unblocked portion of the phase space approaches zero,
$\lambda_{e^-p}^{\rm low}\rightarrow 0$.  The second of the rate contributions
in Eq.\ (\ref{sumrate}) can be approximated as
\begin{widetext}
\begin{eqnarray}\label{nucaprate_1}
\lambda_{e^-p}^{\rm high} & \approx & \Lambda
{\left[1-e^{\xi_{np}-\eta_e+\eta_{\nu_e}}\right]}^{-1}
\int_{\epsilon}^\infty{x^2{\left(x+\xi_{np}\right)}^2 {\left(
{{1}\over{e^{x+\xi_{np}-\eta_{e}}+1}}-{{1}\over{e^{x-\eta_{\nu_e}}+1}} \right)}
dx} \\ & \approx & \Lambda
{\left[1-e^{\xi_{np}-\eta_e+\eta_{\nu_e}}\right]}^{-1} \sum_{n=0}^{4}{\beta_n
\left[F_n\left(\eta_e-\xi_{np}-\epsilon\right)-F_n\left(\eta_{\nu_e}-\epsilon\right)\right]},\nonumber\end{eqnarray}
\end{widetext} where the notation is as above and where the $\beta_n$
are as defined above for Eq.\ (\ref{approxrate1}). In summary, a hole
in the low energy $\nu_e$ distribution results in a lower value for
$\lambda_{\nu_en}$, a higher value for $\lambda_{e^-p}$ and, hence, an
increased $n/p$ ratio.  The rate $\lambda_{e^-p}$ is illustrated in
Fig.~\ref{eleccap2} as a function of $\epsilon$ or $L_{\nu_e}$.

By contrast, conversion up to scaled energy $\bar\epsilon=E_{\bar\nu_e}/T$ of
$\bar\nu_e$'s to sterile neutrinos, $\bar\nu_e\rightarrow\bar\nu_s$, would
result in a {\it lower} value of the neutron-to-proton ratio and, hence, a
lower $^4$He yield. This is because a low energy deficit in the $\bar\nu_e$
distribution would lead to a decreased rate for $\bar\nu_e+p\rightarrow n+e^+$
and, on account of less blocking, an increased rate for the reverse
process. Handling the energy threshold for these reactions is, however,
somewhat more complicated than for $\nu_e$ and $e^-$ capture.

Using much the same notation as above, we can approximate the rate for
$\bar\nu_e+p\rightarrow n+e^+$ as
\begin{widetext}
\begin{eqnarray}\label{nubarcaprate}
\lambda_{\bar\nu_ep} & \approx & \Lambda
{\left[1-e^{\xi_{np}-\eta_e-\eta_{\bar\nu_e}}\right]}^{-1} \int_{\gamma_{\rm
thresh}}^\infty{x^2{\left(x-\xi_{np}\right)}^2 {\left(
{{1}\over{e^{x-\eta_{\bar\nu_e}}+1}}-{{1}\over{e^{x-\xi_{np}+\eta_{e}}+1}}
\right)} dx} \\ & \approx & \Lambda
{\left[1-e^{\xi_{np}-\eta_e-\eta_{\bar\nu_e}}\right]}^{-1}
\sum_{n=0}^{4}{\bar\alpha_n \left[F_n\left(\eta_{\bar\nu}^{\rm
eff}\right)-F_n\left(\eta_{\bar e}^{\rm
eff}\right)\right]}.\nonumber\end{eqnarray}
\end{widetext}
The integration variable in the first of these equations is
$x=E_{\bar\nu_e}/T$, and the final state positron energy will be $E_{e^+}
=T\left(x-\xi_{np}\right)$. The scaled energy threshold in these expressions is
\begin{eqnarray}
\label{thresh}
\gamma_{\rm thresh} & = & \xi_{np}+m_e\ \ \ {\rm for}\ \
\xi_{np}+m_e\ge\bar\epsilon \\ \gamma_{\rm thresh} & = & \bar\epsilon \ \ \ \ \
\ \ \ \ \ \ \ \ {\rm for}\ \ \bar\epsilon>\xi_{np}+m_e\nonumber
\end{eqnarray}
where $m_e\equiv m_ec^2/T$. It is clear that transformation of $\bar\nu_e$'s
with energies below the threshold energy $Q_{np}+m_ec^2$ does not affect the
rate. In the second approximation in Eq.\ (\ref{nubarcaprate}), the effective
$\bar\nu_e$ degeneracy parameter is $\eta_{\bar\nu}^{\rm eff}
=\eta_{\bar\nu_e}-\bar\epsilon$, while the effective positron degeneracy
parameter is $\eta_{\bar e}^{\rm eff} = \xi_{np}-\eta_e-\bar\epsilon$. (Since
electromagnetic equilibrium always obtains here, the positron and electron
degeneracy parameters have equal magnitudes and opposite signs,
$\eta_{e^+}=-\eta_e$.) If we define $a\equiv 2\bar\epsilon-\xi_{np}$ and
$b\equiv \bar\epsilon\left(\bar\epsilon-\xi_{np}\right)$, then the coefficients
$\bar\alpha_n$ are: $\bar\alpha_4 =1$; $\bar\alpha_3=2 a$; $\bar\alpha_2=a^2+2
b$; $\bar\alpha_1=2 a b$; and $\bar\alpha_0=b^2$.

Utilizing the same quantities and notation as in Eq.\ (\ref{nubarcaprate}), the
rate for the reverse process of positron capture, $e^++n\rightarrow
p+\bar\nu_e$, can be written as
\begin{widetext}
\begin{equation}
\label{bigger}
\lambda_{e^+n} \approx  {{\Lambda}\over{1-e^{\eta_e-\xi_{np}+\eta_{\bar\nu_e}}}}
\int_{\gamma_{\rm thresh}}^\infty{x^2{\left(x-\xi_{np}\right)}^2 {\left( {{1}\over{e^{x+\eta_e-\xi_{np}}+1}}-{{1}\over{e^{x-\eta_{\bar\nu_e}}+1}} \right)} dx}
+\Lambda \int_{m_e+\xi_{np}}^{\gamma_{\rm thresh}}{ {{x^2{\left( x-\xi_{np} \right)}^2}\over{e^{x-\xi_{np}+\eta_e}+1  }}dx}.
\end{equation}
\end{widetext}
Again we see that if $\bar\epsilon<\xi_{np}+m_e$, then from Eq.\ (\ref{thresh})
the threshold is $\gamma_{\rm thresh}=\xi_{np}+m_e$ and the neutrino flavor
conversion will have no affect on the rate. In this case, the second term of
Eq.\ (\ref{bigger}) will vanish and the first term will be the rate with no
neutrino conversion.  The full rate expression in Eq.\ (\ref{bigger}) can be
broken up into three parts,
\begin{equation}
\label{threeparts}
\lambda_{e^+n} = \lambda_{e^+n}^{\rm first}+\lambda_{e^+n}^{\rm
snd}+\lambda_{e^+n}^{\rm thrd},
\end{equation}
each of which can be rendered in terms of standard relativistic Fermi
integrals.

Here $\lambda_{e^+n}^{\rm first}$ corresponds to the first integral in Eq.\
(\ref{bigger}). It can be reduced to
\begin{equation}\label{firste}
\lambda_{e^+n}^{\rm first}\approx {{\Lambda}\over{
{1-e^{\eta_e-\xi_{np}+\eta_{\bar\nu_e}}}}} \sum_{n=0}^{4}{\bar\alpha_n
\left[F_n\left(\eta_{\bar e}^{\rm eff}\right)-F_n\left(\eta_{\bar\nu}^{\rm
eff}\right)\right]},\end{equation} where the $\bar\alpha_n$ are as defined for
Eq.\ (\ref{nubarcaprate}), the effective positron degeneracy parameter is
$\eta_{\bar e}^{\rm eff} \equiv -\eta_e+\xi_{np}-\bar\epsilon$, and the
effective $\bar\nu_e$ degeneracy parameter in this case is $\eta_{\bar\nu}^{\rm
eff} \equiv \eta_{\bar\nu_e}-\bar\epsilon$.

Note that the second integral in Eq.\ (\ref{bigger}) is the sum
$\lambda_{e^+n}^{\rm snd}+\lambda_{e^+n}^{\rm thrd}$. The last term in this sum
can be approximated as
\begin{equation}\label{thirde}
\lambda_{e^+n}^{\rm thrd} \approx -\Lambda \sum_{n=0}^{4}{\bar\alpha_n
F_n\left({\xi_{np}-\eta_e-\bar\epsilon}\right)},
\end{equation} 
where the $\bar\alpha_n$ are the same as defined above for Eqs.~(\ref{nubarcaprate}) \& Eq.\ (\ref{firste}).  In similar fashion we can express
$\lambda_{e^+n}^{\rm snd}$ in terms of standard relativistic Fermi integrals,
\begin{equation}\label{seconde}
\lambda_{e^+n}^{\rm snd} \approx \Lambda \sum_{n=0}^{4}{\bar\beta_n F_n\left({-m_e-\eta_e}\right)}.\end{equation}
We define $x\equiv 2 m_e +\xi_{np}$ and $y=m_e\left(m_e+\xi_{np}\right)$, with $m_e\equiv m_ec^2/T$.
With these definitions we can write the $\bar\beta_n$ in Eq.\ (\ref{seconde}) as: $\bar\beta_4\equiv 1$, while  $\bar\beta_3\equiv 2x$, $\bar\beta_2\equiv x^2+2y$,  $\bar\beta_1\equiv 2xy$, and  $\bar\beta_0\equiv y^2$.

\end{document}